\documentclass[trackchanges, twocolumn]{aastex701}

\usepackage{amsmath}
\usepackage{booktabs} 

\usepackage{newtxtext,newtxmath}
\usepackage[T1]{fontenc}
\usepackage{ulem}

\usepackage{graphicx}
\usepackage{amsmath}
\usepackage{xspace}
\usepackage{xcolor}
\usepackage{listings}
\usepackage{lipsum}
\usepackage{threeparttable}
\usepackage{gensymb} 

\usepackage[all]{hypcap}

\begin{document}

\title{NATA: a chemical-picture equation of state for stellar merger hydrodynamics.}

\author[orcid=0000-0001-6251-5315,gname=Natalia,sname=Ivanova]{Natalia Ivanova}
\affiliation{Department of Physics, University of Alberta, Edmonton, T6G 2E7, Alberta, Canada}
\email{nata.ivanova@ualberta.ca}

\begin{abstract}
We present NATA (Nearly Always Truthful Approximations), a self-contained chemical-picture equation of state for hydrogen-helium-carbon-oxygen mixtures, developed for hydrodynamics calculations of stellar mergers and common envelope evolution. 
NATA carries bound species explicitly in ionisation and dissociation equilibrium and evaluates pressure, internal energy, and entropy from one continuous, single-physics chemical-picture solve built on analytic and semi-analytic component physics, rather than by stitching together component EOS tables.
The result is smooth thermodynamics across ionisation, dissociation, and degeneracy fronts, with monotonic internal energy and no artificial energy steps at EOS seams, so the EOS inverts cleanly for codes that advance internal energy.  We benchmark NATA against the blended MESA EOS, ab initio hydrogen-helium tables, and reference-independent analytic limits. NATA reproduces clean analytic limits and selected fully ionised benchmarks at the sub-percent level, while most remaining percent-level residuals are confined to partial-ionisation and dissociation fronts; these residuals are traced to identifiable picture choices, front placements, and reference-table seams or extrapolations.  NATA therefore provides a well-behaved fallback in cool, low-density regimes where blended EOS tables may be stretched beyond their intended use and become physically unreliable.
\end{abstract}


\section{Introduction}
\label{sec:intro}

``Well, nobody's perfect,'' and the same is true of equations of state (EOS).
The imperfection matters most when an EOS is used outside the regime for which it was built.
An EOS sits at the heart of almost everything in astrophysics: understanding nearly any
astronomical object relies on one or another of them.  A great deal of effort has
gone into making each as accurate as possible, but that accuracy comes at a
price.  Most precise, carefully constructed EOS were built with a specific object or an event 
in mind, and they are therefore valid only over the limited range of density and
temperature that the object in question demands.

Stars are one such object, and an awkward one: a single stellar model spans many orders of magnitude in density and
temperature, often beyond the validated domain of any one specialised EOS.  
The well-known remedy, from the {\tt MESA} project
\citep{Paxton2011,Paxton2013,Paxton2015,Paxton2018,Paxton2019,Jermyn2023}, is to
stitch several EOS together across the density-temperature plane.  Because it is
both accurate and readily available, the {\tt MESA} EOS has become widely used,
including in the hydrodynamics codes that model stellar mergers, as in our own
group.  Such codes typically use the EOS in inverted form: the {\tt MESA}
$\rho-T$ table is remapped onto a $\rho-u$ (or $\rho-S$) table, so that pressure
and temperature are recovered from density and internal energy, which is what a
scheme that advects $u$ requires.  To understand the underlying issue, we look
more closely at how such equations of state are constructed.

Equations of state for partially ionised matter are built in one of two main
pictures, distinguished by what is taken as fundamental.  The chemical picture
treats the plasma as a mixture of a prescribed set of species, molecules, atoms,
ions, and free electrons, and fixes their abundances by chemical equilibrium among
them, in practice by minimising a total free energy written as a sum over those
species subject to stoichiometry and charge neutrality.  The bound states are thus
an input and their populations an output, and this is what defines the picture,
independently of whether the words "free energy" appear in a given code's name.
It is the route of the MHD, Mihalas--Hummer--D\"appen,
\citep{HummerMihalas1988,Mihalas1988} SCVH \citep{Saumon1995}, and FreeEOS \citep{Irwin2012} equations of state, which formulate the equilibrium in terms of a species free energy.
Earlier approximate stellar EOSs such as EFF \citep{Eggleton1973} belong to the same
chemical-picture tradition, but with more approximate analytic treatments.
The picture treats ionisation and dissociation transparently and at low cost, but
it requires the species list to be fixed in advance and an explicit prescription
for how bound states dissolve under compression, the occupation-probability or
continuum-lowering term, which is a modelling choice and grows ambiguous across
the pressure-ionisation transition where the notion of a bound state itself breaks
down.

The physical picture instead takes only electrons and nuclei, interacting through
the Coulomb potential, as fundamental, and lets the bound states emerge from the
many-body problem, evaluated by activity expansion, as in OPAL
\citep{RogersNayfonov2002}, or by ab initio quantum simulation, as in the
hydrogen-helium tables of \citet{ChabrierMazevetSoubiran2019,ChabrierDebras2021}
built on the quantum molecular dynamics of \citet{MilitzerHubbard2013}.  
This reduces the species ambiguity and is closer to first-principles where it
applies, but it still depends on the chosen many-body approximation,
exchange-correlation treatment, and interpolation of tabulated results.  It is
also computationally expensive and is delivered only as tables over a bounded
density-temperature domain, so it does not by itself span the full plane that a
stellar-merger calculation visits.

A third and simpler class sidesteps ionisation altogether by assuming the plasma
to be fully ionised: HELM \citep{TimmesSwesty2000}, a Helmholtz free-energy table
for a completely ionised gas of electrons and nuclei plus radiation, is the
standard example, and despite its name it belongs to neither picture, since it
carries no bound states and solves no ionisation balance.  It is accurate where
the matter is in fact fully ionised and the interactions are weak compared
with the kinetic energies, and incorrect where it is not, which is the
source of the wrong fallback discussed below.  Where the coupling is
significant it requires supplementary non-ideal terms, and {\tt MESA}
accordingly hands over to PC or Skye
\citep{PotekhinChabrier2010,Jermyn2021}.

It is clear from the above that the {\tt MESA} EOS combines components drawn
from several approaches: chemical-picture partial-ionisation EOSs,
physical-picture dense-matter EOSs, and the fully ionised limit.  The MESA
framework blends these components smoothly in the forward  $(\rho,T)$  EOS
\citep{Paxton2011,Paxton2019,Jermyn2023}.  However, smooth blending of the
returned quantities is not the same as placing all components on a single
physical energy reference.  Each component carries its own physical assumptions,
tabulated domain, and energy convention.  Residual offsets or changes in slope
across component transitions can therefore become important when the blended EOS
is remapped and inverted in variables such as $(\rho,u)$.

Two difficulties follow.  The first is coverage: the {\tt MESA} tables do not span
the full domain a merger calculation visits, and at low density and low
temperature the blend falls back to fully ionised matter.  In parts of the
plane the underlying tables are not solved in the chemical picture and instead
return the fully ionised HELM result.
In the cool, very low-density part of the plane, below the tabulated coverage of
the partial-ionisation components, the implemented MESA blend falls back to
the fully ionised HELM EOS rather than to a neutral or molecular gas.  
In the cases relevant here, the fallback sets in around \(\log_{10}T\lesssim3\) and \(\log_{10}\rho\lesssim -9\), and gives a fully ionised mean molecular weight and free-electron pressure where the physical gas would be neutral or molecular; its full extent and lower edge, which are not documented and can only be inferred from where SCVH stops and HELM is returned, are given in \S\ref{sec:mesa_sm}. \footnote{See also the {\tt MESA} issue tracker,
\url{https://github.com/MESAHub/mesa/issues/995} \label{fn:mesa995}.}

The second difficulty, foreshadowed by the seams above, is more general: the
stitching itself introduces discontinuities.  HELM is a fully ionised
electron-ion-radiation EOS and omits the ionisation and dissociation energy that
the chemical-picture components include, so the zero point of the internal energy
differs between blend regions.  While {\tt MESA} reconciles these offsets well
enough for forward stellar evolution, which uses only local energy differences and
derivatives, they become observable when the EOS is inverted to obtain temperature
as a function of density and internal energy, $T(\rho, u)$, as required by an SPH (smoothed particle hydrodynamics)
or grid hydrodynamics code that advects $u$.  Crossing a blend seam at fixed
$\rho$ then produces a step or kink in $u$, which on inversion appears as a
spurious jump in temperature and pressure at a contact with no physical origin.  A
table built by inverting the blended {\tt MESA} EOS can therefore introduce
artificial discontinuities into a hydrodynamic calculation.

These obstacles motivate a single-physics equation of state with one consistent
energy reference across the entire plane, covering the full range relevant to
stellar mergers and a little beyond.  That is what we present here: an equation of
state we call NATA, for Nearly Always Truthful Approximations.
We adopt the
chemical picture \citep{Saumon1995,Irwin2012}, because it can be evaluated with
analytic and semi-analytic component physics by a local chemical-equilibrium
solve at any point of the plane, returning a value over the whole
density-temperature domain without table gaps, and therefore without the
component boundaries that motivate this work.
We differ from the codes above in one respect: rather than obtaining
the populations by minimising a global free energy, as MHD, SCVH, and FreeEOS do \citep{HummerMihalas1988,Saumon1995,Irwin2012}, we solve the ionisation and
dissociation equilibrium directly,  without a
global numerical search.  A global minimisation ties the result to a single model
free energy whose non-ideal terms are each valid only in particular regimes, and
to a numerical search that must reconverge at every point and can lose continuity
near pressure ionisation and at extreme degeneracy; guaranteeing a smooth,
convergent solution across the full merger plane is therefore demanding, and the
available free-energy equations of state are correspondingly built and validated
over bounded domains.  
Solving the equilibrium directly, with analytic and semi-analytic
non-ideal corrections, sidesteps the minimisation and returns a continuous,
single-valued result across the whole density-temperature plane.
We do not claim that it is perfect; as with every EOS before it, it is not.  What it provides is the property the present application needs: a continuous chemical-picture thermodynamic surface with a single energy zero point, free of stitching seams, fall-back regions, and the inversion artefacts they produce.

The paper first describes the physics included in the EOS.  It then presents
comparison and validation against the {\tt MESA} EOS and the hydrogen-helium
tables of \citet{ChabrierDebras2021}.  We close by delimiting the region of safe
applicability and explaining how to use the code and the tabulated output.

\section{Equation of state}

\label{sec:nata_eos}

The NATA EOS is evaluated locally and directly at each prescribed pair
$(\rho,T)$,  using the same species-resolved Fermi-Dirac-Saha chemical EOS everywhere.   In this single-closure sense the implementation is seamless across the $\rho-T$ plane. The regions where this closure is accurate, or where it remains unvalidated, are discussed separately in \S\ref{sec:validation}.

A specific composition may be prescribed directly.  In the \texttt{hheco} mode,
the input H, He, C, and O mass fractions are used as given, so any chosen
H/He/C/O mixture can be evaluated.  In the \texttt{gs98} mode \citep{GrevesseSauval1998}, the driver fixes the carbon
and oxygen masses from a GS98-like abundance table of nineteen elements between carbon and nickel (C, N, O, Ne, Na, Mg, Al, Si, P, S, Cl, Ar, K, Ca, Ti, Cr, Mn, Fe, Ni): carbon receives its tabulated mass fraction of $Z$ and the remaining metal mass, the heavier elements folded together, is assigned to oxygen.  Because carbon, oxygen, and the folded metals all have $Z/A\simeq0.5$, this preserves the total metal mass and the electron fraction and gives a charge-consistent reduced H/He/C/O mixture.  The table enters only through this mass split, and the user may replace it with their own element fractions.  It is nonetheless an approximation: the ionisation ladders, the particle counting, and hence $\mu$, all run on the carbon and oxygen proxies carrying their true charge ladders, not on an element-by-element EOS.
If no explicit H/He/C/O values are supplied in \texttt{hheco} mode, the driver
uses the input $X,Y,Z$ and maps the metal mass to oxygen.

The chemical network contains
\[
{\rm H}_2,\ {\rm H},\ {\rm H}^+,\ {\rm H}^-,\ {\rm H}_2^+,
\]
\[
{\rm He},\ {\rm He}^+,\ {\rm He}^{2+},\ {\rm He}_2^+,\ {\rm HeH}^+,
\]
carbon stages $({\rm C},\ldots,{\rm C}^{6+})$, oxygen stages
$({\rm O},\ldots,{\rm O}^{8+})$, and free electrons.  The H/He ionisation
states and the C/O ion ladders are obtained from Saha relations, while the
molecular and molecular-ion abundances are obtained from mass-action relations
\citep[e.g.][]{Mihalas1978,Mihalas1988,LL5}.
For C and O the internal partition function of each ionisation stage is a
truncated sum over the ground term and low-lying excited terms,
$U_i(T)=\sum_k g_k\exp(-E_k/k_{\rm B}T)$, with term energies $E_k$ and
statistical weights $g_k=\sum_J(2J+1)$ taken from the NIST Atomic Spectra
Database \citep{NIST_ASD}.  Stages
whose first excited terms lie at several hundred eV, and the bare nuclei, are
represented by their ground-state statistical weight.

Such a sum cannot simply be cut at the measured terms.  Continued over the
Rydberg series it diverges, because the level degeneracy grows while the
excitation energy saturates at the ionisation limit.  What removes the divergence
is the occupation probability.  For H and He the partition functions are already
occupation-weighted level sums in the Hummer-Mihalas sense: the level volume grows
as $n^{6}$, so the occupation factor falls faster than the Boltzmann factor
saturates and the series converges at any nonzero perturber density.  The
numerical sum is cut at $n=50$, and the cutoff matters less as the density rises,
because the occupation factor then removes the high-$n$ levels at lower $n$.

The C and O measured lists are completed in the same way.  A hydrogenic
Rydberg series converging on each ionisation limit is appended, starting at $n=5$
for the many-electron stages and at $n=2$ for the helium-like and hydrogen-like
stages, with levels below the highest retained measured term skipped so that the
two descriptions do not overlap.  These levels carry the same excluded-volume
dissolution, keyed here to the heavy-particle density, since the mechanism is
hard-sphere overlap.  The measured and completed terms contribute to the returned
internal energy and entropy as well as to the balance, and the density-dependent
Rydberg occupation carries its pressure partner.

The electron degeneracy parameter
$\eta$ is solved from charge neutrality, and the corresponding
finite-temperature Fermi-Dirac electron density is used in the chemical
equilibrium.  Where the charge-neutrality equation has multiple roots in cold
dense pressure-ionising regions, the root search selects the high-$\eta$,
high-ionisation branch, avoiding a cell-to-cell switch between weakly ionised
and pressure-ionised solutions; in the fully neutral limit it instead accepts
the vanishing-electron solution.

This is a numerical branch selection rather than a construction of phase
equilibrium, and the selected surface is not everywhere mechanically stable.  In a
bounded part of the carbon-oxygen pressure-ionisation region, near
$\log\rho\simeq0.35$ to $0.9$ and $\log T\simeq4.4$ to $5.25$, the returned
pressure has $(\partial P/\partial\rho)_T<0$ and the first adiabatic exponent $\Gamma_1<0$.
In a thermodynamically
consistent free-energy EOS such a branch would be replaced by a phase-coexistence
construction.  NATA cannot make that construction, because its chemical solve is
not obtained by minimising one total Helmholtz free energy, so the pressure
inversion is not a prediction of a physical phase boundary.  SCVH provides the
contrast: its plasma phase transition follows from an instability of its
free-energy model and the corresponding coexistence curve \citep{Saumon1995}.  We
therefore treat the multiple roots as a branch-selection problem only, and identify
the pressure-inversion region explicitly as a mechanically unstable part of a
single-phase continuation.

Pressure ionisation is included through a ground-state occupation-probability
correction based on the Hummer-Mihalas formalism
\citep{HummerMihalas1988,Mihalas1988}.  The bound-state survival probability
factorises as $w=w_{\rm neutral}\,w_{\rm charged}$.  The neutral term
$w_{\rm neutral}$ is a hard-sphere excluded-volume dissolution term, keyed to
the fixed parent-nuclei density of the species.  The charged term
$w_{\rm charged}$ is keyed to the free-charge density $n_{\rm pert}$ that
sources the microfield.
It uses the Holtsmark normal-field scaling
$F_0 \propto n_{\rm pert}^{2/3}$ \citep{Holtsmark1919}, while the
characteristic field scale is reduced relative to the bare Holtsmark value by
$g(\kappa)=(1+\kappa)\exp(-\kappa)$, with $\kappa$ the perturber spacing measured
in screening lengths.
The reduction is a numerical regularisation rather than an improved
microfield model, introduced to limit the ionisation feedback through
$n_{\rm pert}$, which the balance itself returns.  Its form is borrowed from the
screened field of a perturber at the ion-sphere radius, not derived from a
screened microfield distribution.

The resulting prescription is therefore not a screened Holtsmark
distribution.  The Holtsmark distribution is the uncorrelated, unscreened
weak-coupling limit, whereas screening and ionic correlations also change the
shape of the distribution, as \citet{PotekhinChabrierGilles2002} show.  We retain
the scale reduction only as a regularisation of that feedback, not as a
pressure-stability prescription.
\footnote{The
\citet{PotekhinChabrierGilles2002} distributions are implemented as a selectable
alternative, mapped onto an effective one-component plasma.  They are not the
production default: the substitution degrades the agreement with
\citet{ChabrierDebras2021} on the H/He plane.}

The \texttt{alpha\_HM\_H2} parameter controls an optional
additional $\mathrm{H_2}$ pressure-dissociation factor, off by default.  The H
and He occupation-weighted excited-state partition-ratio corrections, and their
pressure, energy, and entropy partners, are enabled at full strength in the
current implementation: the corresponding namelist inputs
\texttt{alpha\_HM\_H} and \texttt{alpha\_HM\_He} are read for compatibility, but
the ionisation-control setter overrides them and pins both to unity.  The input
also allows scale factors on the H, He, and He$^+$ ionisation energies,
defaulting to unity; these are diagnostic controls, not separate EOS branches.

Continuum lowering enters the balance of every atomic ionisation ladder included
in the Saha solve as an explicit ionisation-potential depression.  In the
production setup this depression is a degeneracy-aware screened Stewart-Pyatt form
\citep{StewartPyatt1966}, with a screening length obtained from the
finite-temperature Fermi-Dirac electron compressibility, and it
is applied to the H, He, He$^+$, and carbon and oxygen stages alongside the
occupation-probability pressure ionisation, so each ladder is dissolved by both
channels.  The code additionally reconstructs a Coulomb free-energy density
$F_{\rm C}=\rho(u_{\rm C}-Ts_{\rm C})$ from the same nonideal charged-particle
package that returns the Coulomb thermodynamic quantities
\citep{ChabrierPotekhin1998,PotekhinChabrier2000,Groth2017}, and can optionally
replace the H, He, and He$^+$ depressions with the consistent
$-\partial F_{\rm C}/\partial N_i$, finite-differenced over the corresponding
ionisation reactions with the carbon and oxygen populations held in the Coulomb
background; the Stewart-Pyatt value is then the warm start and the fallback where
this is unavailable, and when this mode is active the pressure-positivity guard
factor is fed back into the depression so that the Saha shift and the returned
Coulomb correction use the same guarded $F_{\rm C}$.  This $F_{\rm C}$-consistent
mode is thermodynamically tidier but makes pressure ionisation a positive-feedback
loop with multi-root behaviour, so it is off by default and is not used for the
reported tables, and the carbon and oxygen stages never receive a per-stage
$F_{\rm C}$ depression.  This remains an approximation: the correction is derived
from the reconstructed Coulomb free energy of the adopted component fits, while
the full EOS is still a fixed-point chemical-picture solve rather than a
variational minimisation of one total Helmholtz free energy over all species.

Molecular hydrogen uses an internal partition function built as a direct sum over the bound rovibrational levels of the $X\,^1\Sigma_g^+$ ground state.  The level term values are taken from the $\mathrm{H_2}$ Dunham constants
\citep{HuberHerzberg}, referenced to $v=0,J=0$, with the homonuclear symmetry
factor $\sigma=2$; ortho- and para-$\mathrm{H_2}$ nuclear-spin states are not
resolved separately.  The same $Q_{\mathrm{H_2}}(T)=\sigma^{-1}\sum_{v,J}(2J+1)
\exp[-E_{v,J}/k_{\rm B}T]$ sets the $\mathrm{H_2}$ dissociation constant and the molecular internal energy and entropy, so the dissociation balance and the energy it carries derive from a single function.  This direct sum follows the method of \citet{BarklemCollet2016} and captures the vibrational anharmonicity and the $v$-dependent rotational constant $B_v=B_e-\alpha_e(v+\tfrac12)$ that shape the level density near dissociation, where a rigid-rotor plus harmonic-oscillator form would undercount the bound levels.  The molecular ions $\mathrm{H}_2^+$, $\mathrm{He}_2^+$, and $\mathrm{HeH}^+$ are included approximately through ground-state dissociation energies and simple statistical factors only, not through full molecular partition functions.

The H$_2$ dissociation energy is taken to be density dependent rather than fixed.
We use the saturating molecular continuum-lowering form
\[
D_{\mathrm{H}_2}^{\rm eff}(\rho)=\frac{D_{\mathrm{H}_2}}
{1+(\rho/\rho_{\mathrm{diss}})^{p}},
\]
which weakens the bond as the density rises and gives molecular dissociation
$\mathrm{H}_2\rightarrow 2\,\mathrm{H}$ its own density scale, separate from and
below the atomic pressure ionisation that acts through the H ionisation-energy
depression.  In all tables presented here we adopt
$\rho_{\mathrm{diss}}=0.8\,{\rm g\,cm^{-3}}$, set to the molecular-to-atomic
transition scale of the hydrogen phase diagram
\citep{ChabrierMazevetSoubiran2019,ChabrierDebras2021}, and $p=4$.  These values
are fixed in advance and are held the same across all compositions and reference
comparisons.  The effective bond enters the H$_2$ mass-action constant, so the
composition is always solved with the density-dependent dissociation energy.  It
also carries the mechanical partner
\[
P_{\mathrm{H}_2,\rm bind}=-\rho\,n_{\mathrm{H}_2}\,
\frac{\mathrm{d}D_{\mathrm{H}_2}^{\rm eff}}{\mathrm{d}\rho},
\]
so the pressure contribution follows from the same effective bond prescription and
vanishes where no H$_2$ remains.  The same $D_{\mathrm{H}_2}^{\rm eff}$ sets the
molecular term of the binding-energy reference for $u$ when the binding energy is
reported, while the composition shift and pressure partner are present regardless.

The nonideal charged-particle contribution is evaluated as one combined package.
The classical ion term is the liquid one-component-plasma reduced Coulomb free
energy of \citet{PotekhinChabrier2000} with linear mixing over the ion species,
each species coupling at $\Gamma_j=Z_j^{5/3}\Gamma_e$ on the shared electron
background \citep{ChabrierPotekhin1998}.  The fit runs smoothly from the
Debye-Huckel limit $f\rightarrow-\Gamma^{3/2}/\sqrt{3}$ at weak coupling to the
Madelung linear behaviour at large $\Gamma$, so a single liquid fit is used at all
couplings with no melting cutoff.  The carbon and oxygen ions enter with their
actual charge states, so ${\rm C}^{i+}$ contributes charge $i$ and ${\rm O}^{j+}$
contributes charge $j$, and the index $j$ runs over individual ionisation
  stages rather than over chemical elements.

The coupling parameters are
\[
\Gamma_e=\frac{e^2}{a_e k_{\rm B}T},
\qquad
a_e=\left(\frac{3}{4\pi n_e}\right)^{1/3},
\]
where $\Gamma_e$ is the electron coupling parameter and $a_e$ is the
electron-sphere radius determined by the free-electron density $n_e$. For the
maps below, we also report the population-weighted mean ionic coupling of the
mixture,
\[
\bar{\Gamma}
=
\frac{\sum_j n_j Z_j^{5/3}}{\sum_j n_j}\,\Gamma_e,
\]
where the sums run over the charged ionisation stages. Because both $n_e$ and
the stage populations are outputs of the composition solve, $\bar{\Gamma}$
falls towards zero as the gas recombines and reduces to the usual fully ionised
mixture value when all ionisation ladders are fully stripped.

In addition, the ion quantum correction is taken from the quantum-liquid fit
of \citet{BaikoChugunov2022}, evaluated for each ion species from the ion quantum
parameter
\[
\theta=\frac{\hbar\omega_p}{k_{\rm B}T},
\qquad
\omega_p^2=\frac{4\pi n_{\rm OCP} Z_i^2e^2}{A_i m_{\rm u}},
\]
and the ion density, where $n_{\rm OCP}=n_e/Z_i$ is the equivalent
one-component-plasma density on the shared electron background.  The quantum
energy, pressure, and entropy follow from the same fitted free energy.  The fit
reproduces the leading Wigner-Kirkwood behaviour
\citep{Wigner1932,Kirkwood1933} at small $\theta$ and the zero-point scaling at
large $\theta$, so no saturation is imposed.  For the multicomponent mixture the
per-ion correction is weighted by the actual abundance of each species.

Because the same liquid fit is used at all couplings, the Madelung energy and its
pressure are carried by the liquid fit's Madelung asymptote at high $\Gamma$
rather than by a true solid Madelung term, and no liquid-to-solid blend is applied
to the internal energy or the pressure.
The strong-coupling treatment acts on the
entropy only.  The liquid excess entropy $s_{\rm ex}/Nk=\Gamma f'-f$ diverges
negative when extrapolated past melting, so it is ramped smoothly to zero across
the melting window $150\lesssim\Gamma\lesssim200$, leaving the bounded Fermi-Dirac
translational entropy as the ion-entropy floor there.
The quantum entropy is ramped in the same window, while the quantum energy
and pressure are retained.  No separate rolloff is applied as the ion degeneracy
parameter $\theta_i=n_i\lambda_i^3$ crosses unity: the classical translational
entropy, the classical excess entropy, and the quantum correction are instead
carried together, and their cancellation replaces it.
The omitted solid
phonon entropy is small and positive and vanishes as $T\rightarrow0$; adding it
explicitly is a deferred refinement.  This is a controlled regularisation of the
cold dense corner, not a quantum Coulomb-solid EOS or a complete multicomponent
crystallisation treatment.  The crystal branch of \citet{BaikoChugunov2022}
is the appropriate extension, and would require a solid phase and, for this
mixture, a multicomponent phase construction.

The electron contribution to the returned thermodynamics is computed after
the composition solve, using the free-electron density from the charge-neutrality root.
For electrons only, the pressure and kinetic internal energy are evaluated from the
finite-temperature relativistic Fermi-Dirac integrals by direct Gauss-Legendre quadrature,
with the integration interval split at the Fermi surface in the partially and strongly
degenerate regimes, and the electron degeneracy parameter $\eta$ obtained by bisection
on the same quadrature; in the strongly degenerate low-temperature limit, where the
finite-temperature quadrature becomes a numerically singular step-function integral,
the analytic zero-temperature relativistic expressions \citep{Chandrasekhar1939}
are substituted instead.  If the pair option is enabled in \texttt{Nata\_eos.input},
the lepton thermodynamics is instead computed from relativistic electron and positron Fermi-Dirac integrals with
\[
n_{e^-}-n_{e^+}=\rho Y_e/m_{\rm u},
\]
including electron and positron pressure, kinetic internal energy, and entropy.
These quantities are evaluated here by direct integration rather than from a
tabulated Helmholtz free energy; cf. the standard relativistic
electron-positron EOS of \citet{TimmesSwesty2000} and the EOS comparison of
\citet{TimmesArnett1999}.  A second input switch controls whether the rest-mass
creation term $2n_{e^+}m_ec^2/\rho$ is included in the reported specific
internal energy.  
The Saha composition solve itself still uses the electron-only
Fermi-Dirac density in the chemical equilibrium; the pair option modifies only the returned lepton thermodynamics at the net charge density.  The pair option is therefore included for diagnostic continuity at high temperature, not as a full pair-equilibrium chemical EOS: in the pair-dominated, baryon-poor regime
($\log T\gtrsim9$, low $\log\rho$) the pairs are not fed back into the composition solve, so NATA is not intended as a pair-equilibrium EOS there, and the returned $\mu$ in that wedge is unreliable (\S\ref{sec:validation}, Table~\ref{tab:reliability}).

The electron exchange-correlation term uses the finite-temperature
uniform-electron-gas fit of \citet{Groth2017}, evaluated on the same positive
atomic-ion charge background used by the OCP linear-mixing package.  Ion-electron
screening is included following \citet{PotekhinChabrier2000}.  The molecular
ions are included in charge neutrality and particle counting, but are not
included in the OCP (one-component plasma) Coulomb mixture, the ion-electron screening sum, or the
positive charge background used by the nonideal charged-particle package.  This
is an approximation restricted to the molecular regime, where their abundance is
expected to be small.

The total pressure is
\[
P = P_{\rm heavy}+P_{\rm lep}+P_{\rm rad}+P_{\rm exc}+P_{\rm nonideal}
+P_{\mathrm{H}_2,\rm bind},
\]
where $P_{\rm heavy}$ is the ideal translational pressure of all non-electron
material species, $P_{\rm lep}$ is the electron or electron-positron Fermi-Dirac
pressure, $P_{\rm rad}=aT^4/3$,
$P_{\rm exc}$ is the excited-state occupation-probability pressure,
which is nonzero in the reported tables because the H, He, and
$\mathrm{H_2}$ partition-ratio corrections that source it are at full strength,
$P_{\mathrm{H}_2,\rm bind}$ is the mechanical partner of the
density-dependent H$_2$ dissociation energy introduced above,
and $P_{\rm nonideal}$ contains the ion Coulomb terms: the liquid OCP branch,
the ion quantum-liquid correction, the electron exchange-correlation, and the ion-electron screening.  A smooth
pressure-positivity guard is retained in the strongly coupled cold-dense corner.
Where this guard activates, the returned $P$, $u$, and $S$ are numerically
regularised rather than exact derivatives of a single Helmholtz free energy.

The specific internal energy is
\[
u = u_{\rm trans}+u_{\mathrm{H_2,int}}
+u_{\rm bind}+u_{\rm exc}+u_{\rm lep}+u_{\rm rad}+u_{\rm nonideal}.
\]
Here $u_{\rm trans}$ is the ideal translational energy of the non-electron
material species, $u_{\mathrm{H_2,int}}$ the H$_2$ internal rovibrational
contribution obtained from the level-sum partition function $Q_{\mathrm{H_2}}$,
$u_{\rm exc}$ the excited-state occupation-probability contribution,  $u_{\rm lep}$ the electron or
electron-positron Fermi-Dirac kinetic energy, which also carries the positron
rest-mass term $2\,n_{e^+}m_ec^2/\rho$ when pairs and the rest-mass switch are on,
and $u_{\rm nonideal}$ the same guarded nonideal package used in the pressure.
With \texttt{include\_binding\_energy = .true.}, the zero is the
molecular-hydrogen reference: residual H$_2$ carries
$(D_{{\rm H}_2}-D_{{\rm H}_2}^{\rm eff}(\rho))$ per molecule from the
density-dependent bond, small and positive under compression and vanishing at low
density; dissociated H carries $D_{{\rm H}_2}/2$ per nucleus; and ionised hydrogen
species carry the corresponding ionisation or molecular-ion energies.  Helium,
carbon, and oxygen are referenced to their neutral atomic ground states,
following the SCVH convention \citep{Saumon1995}.

The entropy is evaluated species by species.  The translational entropy of each
non-electron material species is evaluated with a bounded ideal-gas (Fermi-Dirac)
form,
\[
s_i=n_i k_{\rm B}\left[\frac{5}{2}\frac{F_{5/2}(\eta_i)}
{F_{3/2}(\eta_i)}-\eta_i\right],
\]
where $\eta_i$ is determined from $n_i\lambda_i^3/g_i=F_{3/2}(\eta_i)$.  The
half-integer functions $F_{3/2}$ and $F_{5/2}$ here, and the inverse $F_{1/2}$
that sets the electron-degeneracy correction to the Saha ratios, are evaluated
piecewise rather than from a single series: a non-degenerate fugacity series at
low $\eta$, a Sommerfeld asymptotic expansion at high $\eta$, and Gauss-Legendre
quadrature through the crossover (cubic interpolation on a precomputed monotonic
table for the $F_{1/2}$ inverse), so no series is used outside its convergence
radius.  This reduces to the classical Sackur-Tetrode form in the non-degenerate
limit \citep{Sackur1912,Tetrode1912}. For the nuclei and molecules the Fermi-Dirac
functional is used as a bounded-entropy regularisation that avoids negative
translational entropies at high density, not as a physical fermion or boson
distinction.  The lepton entropy is computed from the relativistic Fermi-Dirac
expression, from the same integrals as the lepton pressure and energy, using the
electron-positron form when the pair option is enabled.  The H$_2$ internal rovibrational
entropy from the same $Q_{\mathrm{H_2}}$, the radiation entropy, the excited-state entropy, and the nonideal charged-particle entropy corrections are
then added to obtain the returned specific entropy.

The reported mean molecular weight $\mu$ is a composition-based particle-counting
quantity.  It is defined from the number density of non-electron material species
plus the free-electron density obtained from the charge-neutrality root,
\[
\mu = \frac{\rho}{m_{\rm u}\left(n_{\rm mat}+n_e\right)} ,
\]
where $n_{\rm mat}$ includes atoms, ions, molecules, and molecular ions, and $n_e$
is the free-electron density used in the Saha solution.  Thus $\mu$ is not obtained
by inverting the total pressure as $P=\rho k_{\rm B}T/(\mu m_{\rm u})$.  Radiation
pressure, electron degeneracy pressure, Coulomb terms, and, when enabled,
electron-positron pair thermodynamics are not folded into this definition.
Thermally produced electron-positron pairs enter the lepton thermodynamics when
the pair option is enabled, but this $\mu$ counts only the net charge-neutrality
free-electron density $n_e$, so both the pair positrons and the additional pair
electrons are excluded.

The EOS returns $P$, $S$, $u$, and the mean molecular weight $\mu$ directly at the
input $(\rho,T)$. When derivative output is enabled, the driver obtains
$c_V$, $c_P$, $\chi_\rho$, $\chi_T$, $\Gamma_1$, $\Gamma_2$, $\Gamma_3$,
$\nabla_{\rm ad}$, and the adiabatic sound speed from local derivatives of the EOS
taken in logarithmic variables.
Here $c_V=(\partial u/\partial T)_\rho$ is the specific heat at constant
volume, and $c_P$ is the specific heat at constant pressure, both per unit mass,
evaluated as $c_P=c_V\Gamma_1/\chi_\rho$;
$\chi_\rho=(\partial\ln P/\partial\ln\rho)_T$ and
$\chi_T=(\partial\ln P/\partial\ln T)_\rho$ are the logarithmic pressure
derivatives; $\Gamma_1$, $\Gamma_2$, and $\Gamma_3$ are the first, second, and
third adiabatic exponents in the notation of \citet{Chandrasekhar1939};
$\nabla_{\rm ad}=(\partial\ln T/\partial\ln P)_S$; and the adiabatic sound speed
is $c_s=(\Gamma_1 P/\rho)^{1/2}$.
Each first derivative is obtained from a local
quadratic least-squares, or Savitzky-Golay, slope on a short symmetric stencil: a
five-point window by default, widened to seven points when the five-point fit is
rough, and falling back to a three-point central slope only when the one-sided
slopes agree. The differencing step is tried first at the base value and then at
successively doubled values; for each requested derivative quantity the first value
that passes the fit-quality and physical-validity tests is retained. This
suppresses round-off scatter at very small steps while keeping the accepted stencil
local. The stencil evaluates neighbouring values of $P(\rho,T)$ and $u(\rho,T)$;
$\chi_T$ and $c_V$ are formed from the $\log T$ derivatives of $P$ and $u$ at fixed
$\rho$, and $\chi_\rho$ from the $\log\rho$ derivative of $P$ at fixed $T$.

Where the central state is fully ionised according to the neutral-fraction
diagnostic, so that the composition response is negligible, $\chi_\rho$ and
$\chi_T$ are instead formed from a semi-analytic component-pressure split,
$\chi = \sum_i (P_i/P)\,\chi_i$. 
The ideal material pressure $P_{\rm heavy}$ contributes
$\chi_\rho=\chi_T=1$, radiation $P_{\rm rad}$ contributes $\chi_\rho=0$ and
$\chi_T=4$, and the electron-positron, Coulomb, excited-state, and molecular-bond pressures $P_{\rm lep}$, $P_{\rm nonideal}$, $P_{\rm exc}$, and $P_{\mathrm{H}_2,\rm bind}$ are carried together as a cleanly differenced residual pressure term.
This avoids the cancellation that a finite difference of the total pressure would otherwise suffer  in radiation- and pair-dominated regimes, where a small density-dependent pressure  contribution can be buried under a much larger nearly density-independent total pressure.
The switch from the finite-difference derivatives to this component split is a
replacement rather than a blend, but it is applied only where the bound fraction
is below one percent, so the composition-response term that the split omits is at
most a percent-level contribution there; the two estimates agree to that order at
the boundary and the $\chi_\rho$ and $\chi_T$ maps show no seam across it.

Across the stencil, the iterative $F_{\rm C}$-consistent continuum-lowering
depression, when that mode is active, is held fixed at its converged centre value
rather than re-iterated at each neighbour. This keeps the stencil free of the
fixed-point convergence noise that the finite difference would otherwise amplify;
in the production Stewart-Pyatt setup the depression is analytic and this freeze is inert. This approximation is applied only over the small derivative stencil. A failed cell is flagged by nonzero \texttt{ierr} and sentinel values, and a  requested derivative is likewise left at the sentinel if the neighbours required by the accepted stencil fail or if the resulting thermodynamic quantity is undefined.

Hydrodynamics codes that advance the internal energy require the equation of
state in the $(\rho,u)$ plane, so we provide a code that builds a companion table on a uniform
$(\log\rho,\log u)$ grid.  It is built by inverting the same NATA EOS used
throughout: at each grid point the temperature is found such that the specific
internal energy returned at $(\rho,T)$ equals the target $u$, and the converged
state supplies the same default quantities as the $(\rho,T)$ table, the
temperature $T$, mean molecular weight $\mu$, total pressure $P$, and specific
entropy $S$, with the finite difference derivative columns appended only when
requested in \texttt{Nata\_eos.input}.  Because every quantity comes from one
inversion of the production EOS, the table is internally consistent with the
$(\rho,T)$ EOS a code would otherwise call at runtime.  The inversion is a scan
of $\log T$ at fixed density to bracket the root of $u(T)-u_{\rm target}$
followed by bisection; convergence is on the energy residual, $u$ matched to
$10^{-8}$ dex, which sets the temperature consistency of the table, with the
bracket width carried only as a fallback guard.  Across the ionisation, dissociation, and recombination fronts $u(\rho,T)$
increases monotonically with temperature at fixed density, with positive specific
heat throughout, so the bracketing scan returns a single root and the inversion is
single valued there by construction.
In the strongly coupled cold dense corner and at high compression, where the
EOS is numerically regularised, the specific heat is not everywhere positive and
$u$ is not everywhere monotonic in temperature.  Where that occurs the root
search selects the physically stable root, the one of positive specific heat at
which $u$ increases through the target, so the tabulated $T(\rho,u)$ remains
single valued and free of the spurious temperature jumps that an arbitrary root
choice would introduce on inversion.

The code is available in a \href{https://doi.org/10.5281/zenodo.20914774}{Zenodo} repository.

\section{Results, Applicability and Validation}

\label{sec:validation}

\begin{figure}
  \centering
  \includegraphics[width=\linewidth]{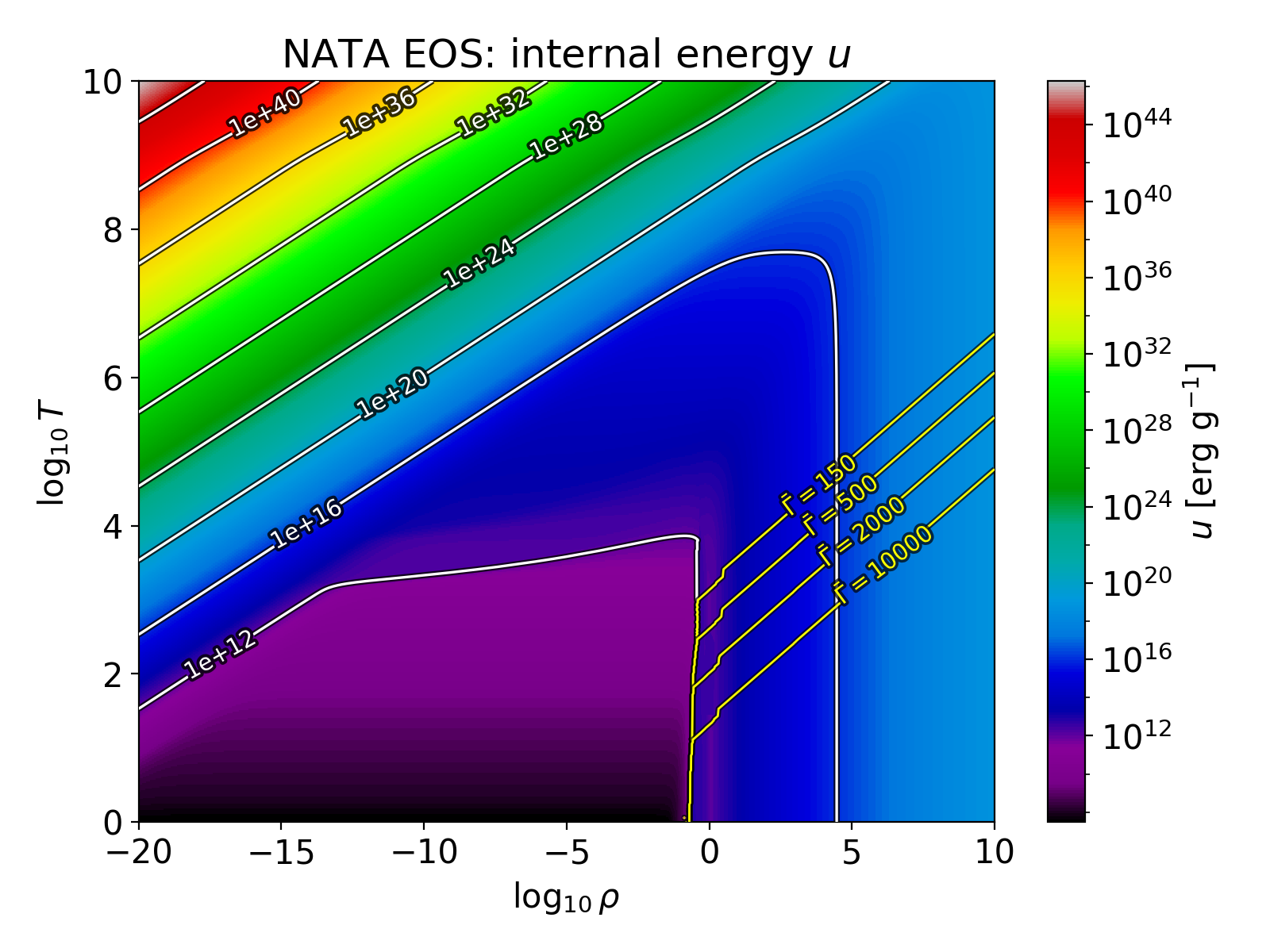}
  
  \includegraphics[width=\linewidth]{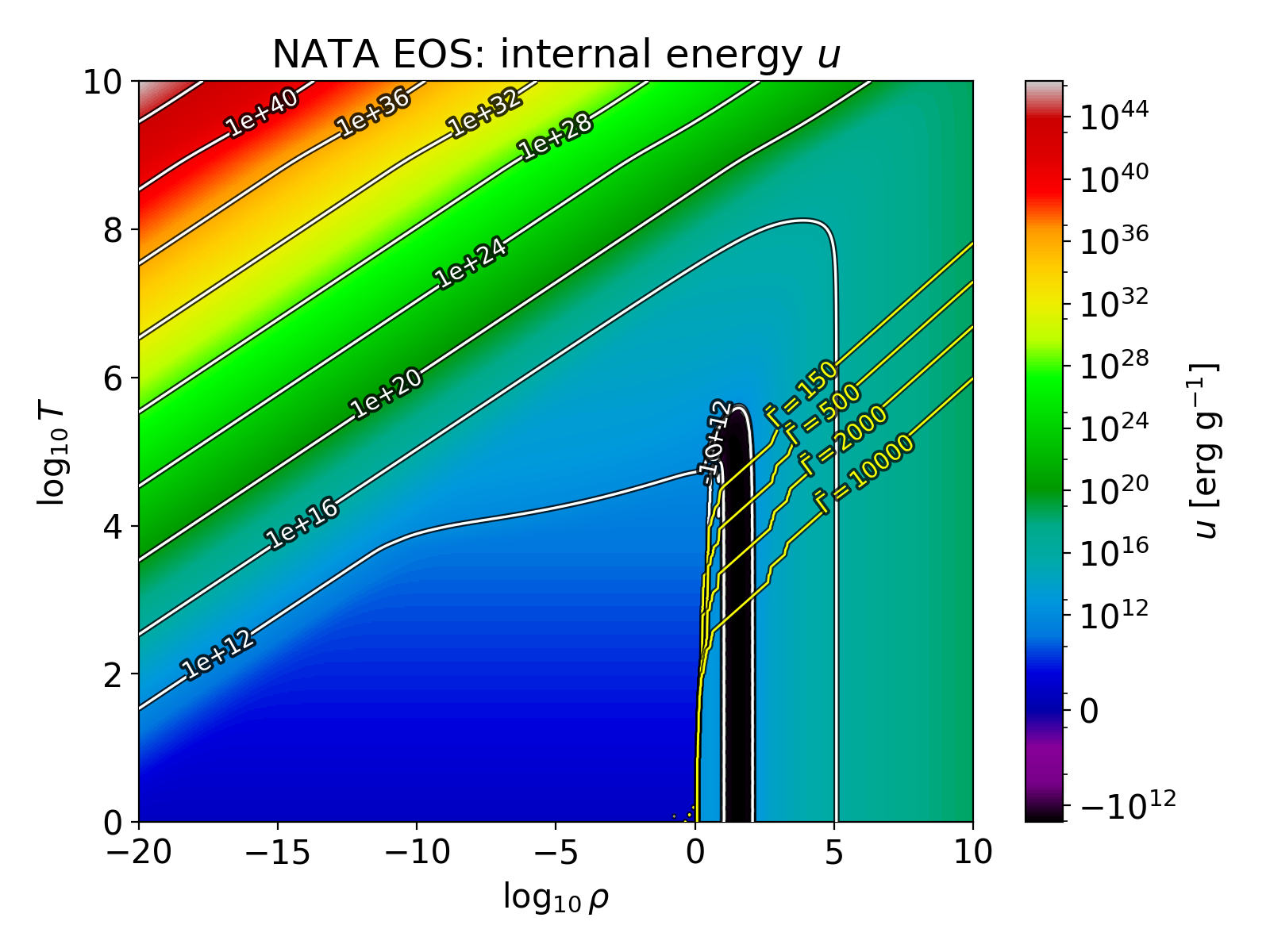}

    \caption{Specific internal energy $u$ of the NATA EOS across the $(\rho,T)$
      plane for the two reference compositions: the stellar mixture with $X=0.7$, $Y=0.28$, and $Z=0.02$ (upper panel),
      and the carbon-oxygen mixture with $C=O=0.5$ (lower panel).
      White lines are contours of constant $u$ (solid for positive values and dashed in
      the negative-energy region of the carbon-oxygen panel),
      while the yellow lines mark the population-weighted mean ionic Coulomb coupling $\bar{\Gamma}$ defined in \S\ref{sec:nata_eos}, evaluated from the solved free-electron density and ion-stage populations. For the carbon-oxygen composition, these contours terminate near $\log\rho\simeq0.5$, where recombination lowers the free-charge density and $\bar{\Gamma}$ falls below the plotted levels.}
\label{fig:u_maps}
\end{figure}

The EOS returns non-failed values over a very large domain,
$-20 \le \log_{10}\rho \le 10$ and $0 \le \log_{10}T \le 10$, for both a stellar
mixture ($X=0.7$, $Y=0.28$, $Z=0.02$, GS98 abundances) and carbon-oxygen material
(pure metal, $Z=1$, $C=O=0.5$).  Other chemical compositions are in principle
possible as input, but here we analyse the results for these two cases.
This is the domain over which the thermodynamics can be evaluated for a
prescribed composition, not a domain over which that composition persists.  NATA
carries no nuclear reactions and does not evolve the abundances, so at the highest
temperatures and densities, where thermonuclear or pycnonuclear burning would
alter the mixture, the tabulated values should be read as numerical coverage and
margin rather than as the state of real matter.
Figure~\ref{fig:u_maps} shows the internal energy across this domain for the two
compositions\footnote{The data tables for these figures, together with the
scripts and figures for all other thermodynamic quantities, including the
derivatives, are available in a \href{https://doi.org/10.5281/zenodo.20914774}{Zenodo} deposit.}
The values are smooth and free
of spurious features over almost the whole plane, which matters for the
validation that follows.

In the carbon-oxygen panel $u$ falls below zero over a bounded region,
$\log\rho\simeq0.65$ to $2.0$ and $\log T\lesssim5.5$.  With the energy referenced
to the neutral C and O ground states there is no positive dissociation baseline,
and the ionisation energy, which is positive, is outweighed there by the negative
ion Coulomb (Madelung) energy.  The sign change is not driven by the ionisation
state: at $\log T=3$ the mean ion charge stays at $4.857$ from $\log\rho=0.9$ to
$2.6$ while $u$ changes sign twice.  The density dependence is set instead by the
Coulomb term and, at higher density, by the growing electron-degeneracy energy.

The ability to return a value should not be read as equal physical reliability everywhere.  The least secure region is the cold, dense, strongly coupled corner, in particular $\log_{10}T \lesssim 2.5$ with the ion coupling parameter at or above the one-component-plasma melting range, $150 \lesssim \Gamma \lesssim 200$.
There the EOS uses the liquid OCP fit at all couplings, with the strong-coupling entropy and the pressure ionisation smoothly regularised, rather than a full
quantum Coulomb-solid or molecular pressure-ionised treatment.  Two effects make
the thermodynamic derivatives least reliable here.  First, the thermal part of $u$
that sets $c_V$ is reduced to the level of double-precision round-off, so $c_V$ and
the quantities built on it carry more residual noise.
Second, because the liquid entropy correction is
ramped to zero across this window while the corresponding energy and pressure
are retained, the returned $S$ is no longer the exact entropy derivative of the
same free energy that yields $P$ and $u$,
so the Maxwell relations are not enforced
there; this is a distinct approximation from the pressure-positivity guard
described in \S\ref{sec:nata_eos}, which acts only where it fires, whereas the
entropy ramp is always active across the melting window.  The adiabatic quantities
$\Gamma_1$, $\Gamma_3$, $\nabla_{\rm ad}$, and the sound speed combine
$c_V=T(\partial S/\partial T)_\rho$ with the $P$ derivatives, so they inherit both
effects.  We therefore regard the results in this corner, energies, pressures, and
derivatives alike, as numerically regularised and useful for continuity tests, not
as validated physical predictions.

A second corner is the hot, rarefied wedge, $\log_{10}T \gtrsim 9$ and
$\log_{10}\rho \lesssim -17$, where the matter is relativistic and
electron-positron pairs are abundant.  The reported $\mu$ remains the
composition-based particle count defined above: it includes the free-electron
density of the electron-only Saha solution but excludes both pair positrons and
the additional pair electrons.  Because the pair populations are not fed back
into the composition solve, this $\mu$ is not a physical total-particle mean
molecular weight once pairs dominate the lepton thermodynamics, and is treated
as diagnostic only in this wedge.

We validate the NATA EOS at three levels: against analytic limits, against the
\texttt{MESA} \texttt{eos} module, and against the \citet{ChabrierDebras2021}
hydrogen-helium tables.

\subsection{Reference-independent checks}
\label{sec:ref_independent}

\subsubsection{Degenerate relativistic electron gas.}
This test checks the degenerate electron limit and its connection to the finite
temperature electron EOS.  In the strongly degenerate, low temperature regime the
NATA electron routine uses the analytic fully degenerate relativistic electron gas
of \citet{Chandrasekhar1939} directly, $P=(\pi m_e^4 c^5/3h^3)\,f(x)$ with
$f(x)=x(2x^2-3)\sqrt{1+x^2}+3\,\mathrm{arcsinh}\,x$ and $x=p_F/m_ec$.  The check
covers the branch selection, constants, units, and the pressure and kinetic energy
normalisation in the EOS driver.  It also confirms that the finite temperature
Fermi-Dirac quadrature approaches this same zero temperature limit as the
degeneracy increases, and that the FD-Saha partial ionisation solve reaches the
same fully ionised degenerate electron pressure.

\subsubsection{Gas plus radiation adiabatic limit.}
In the radiation dominated corner the NATA EOS recovers the exact ideal gas plus
radiation result of \citet{Chandrasekhar1939}, $\Gamma_1=\beta+(4-3\beta)^2
(\gamma-1)/[\beta+12(\gamma-1)(1-\beta)]$ with $\gamma=5/3$, whose radiation limit
is $\Gamma_1\to4/3$ and $\nabla_{\rm ad}\to1/4$.  NATA agrees with $\Gamma_1(\beta)$
at the local $\beta=P_{\rm gas}/P_{\rm tot}$ to five significant figures across the
whole transition from gas to radiation domination.

\subsubsection{One-component-plasma Coulomb term.}
In the liquid OCP branch the NATA ion Coulomb term implements the
\citet{PotekhinChabrier2000} fit (their Eq.~16, in the
\citealp{ChabrierPotekhin1998} lineage).  Driving the live routine over
$-8\le\log\rho\le4$ at $\log T=3.8$, $4.6$, and $5.4$ for pure H, He, C, and O,
NATA tracks PC00 to under $0.04$ percent everywhere and CP98 to about $0.36$
percent, the latter being the genuine gap between the two published fits rather
than a code issue.\footnote{Comparison plots with per element residuals for H,
He, C, and O at the three isotherms, together with the tabulated fits, are
deposited at \href{https://doi.org/10.5281/zenodo.20914774}{Zenodo}.}  This
agreement is set by the coupling rather than by tuning: the reduced free energy
$f(\Gamma)$ depends on $\Gamma$ alone, so every species samples one universal
curve and the residuals differ only in where the density range maps onto it.  At
strong coupling $f(\Gamma)\to A_1\Gamma$ with the published Madelung coefficient
$A_1\simeq-0.9$, and the excess pressure follows the Coulomb virial relation
$P_{\rm ex}/(n_i k_{\rm B}T)=\tfrac{1}{3}\Gamma f'(\Gamma)$, equivalently
$u_{ii}=3P_{\rm ex}/(n_i k_{\rm B}T)$, which NATA reproduces from the
Debye-H\"uckel limit through to the melting value $\Gamma\simeq175$.


\begin{figure*}
  \gridline{\fig{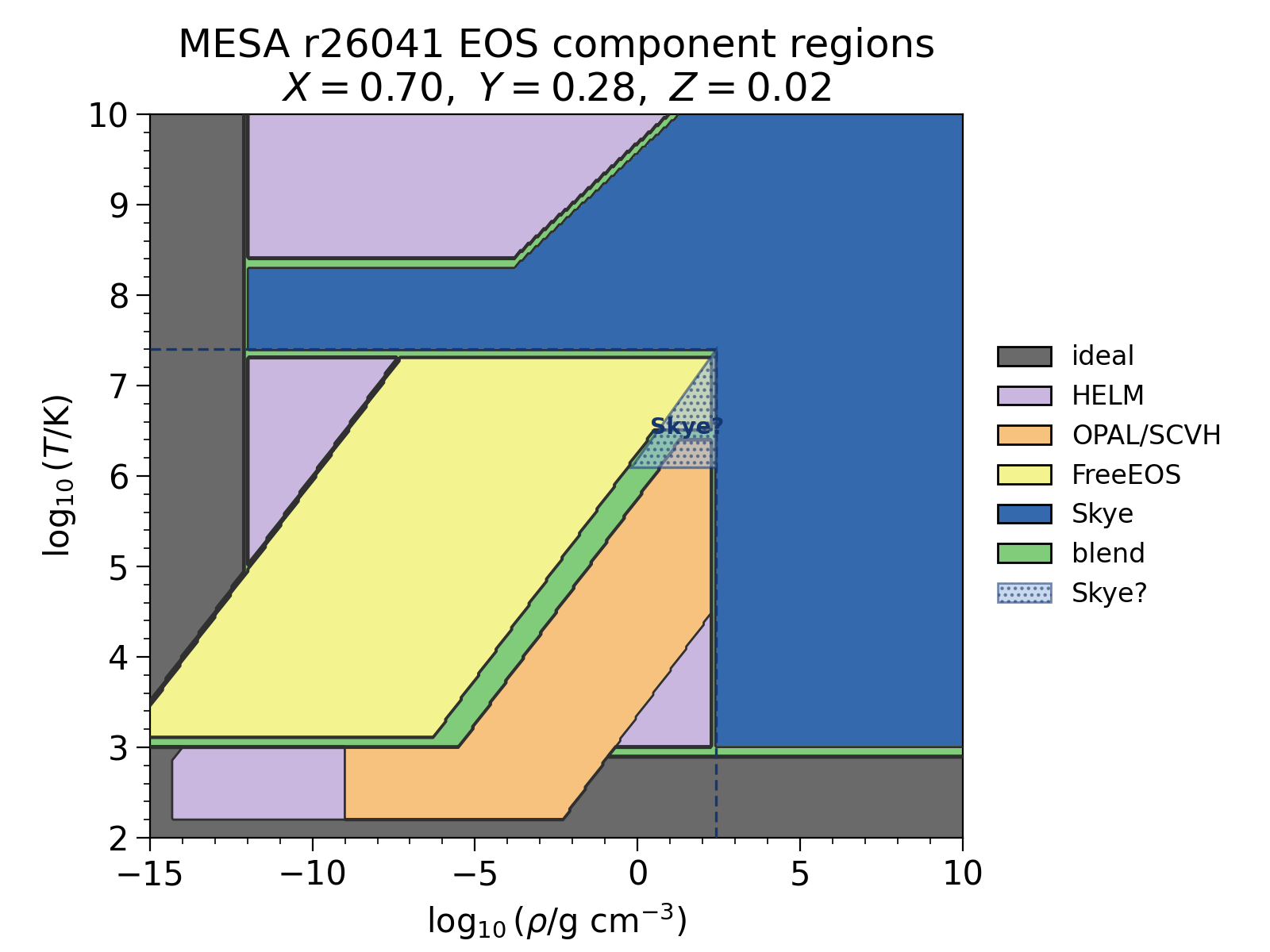}{0.49\textwidth}{(a)}
            \fig{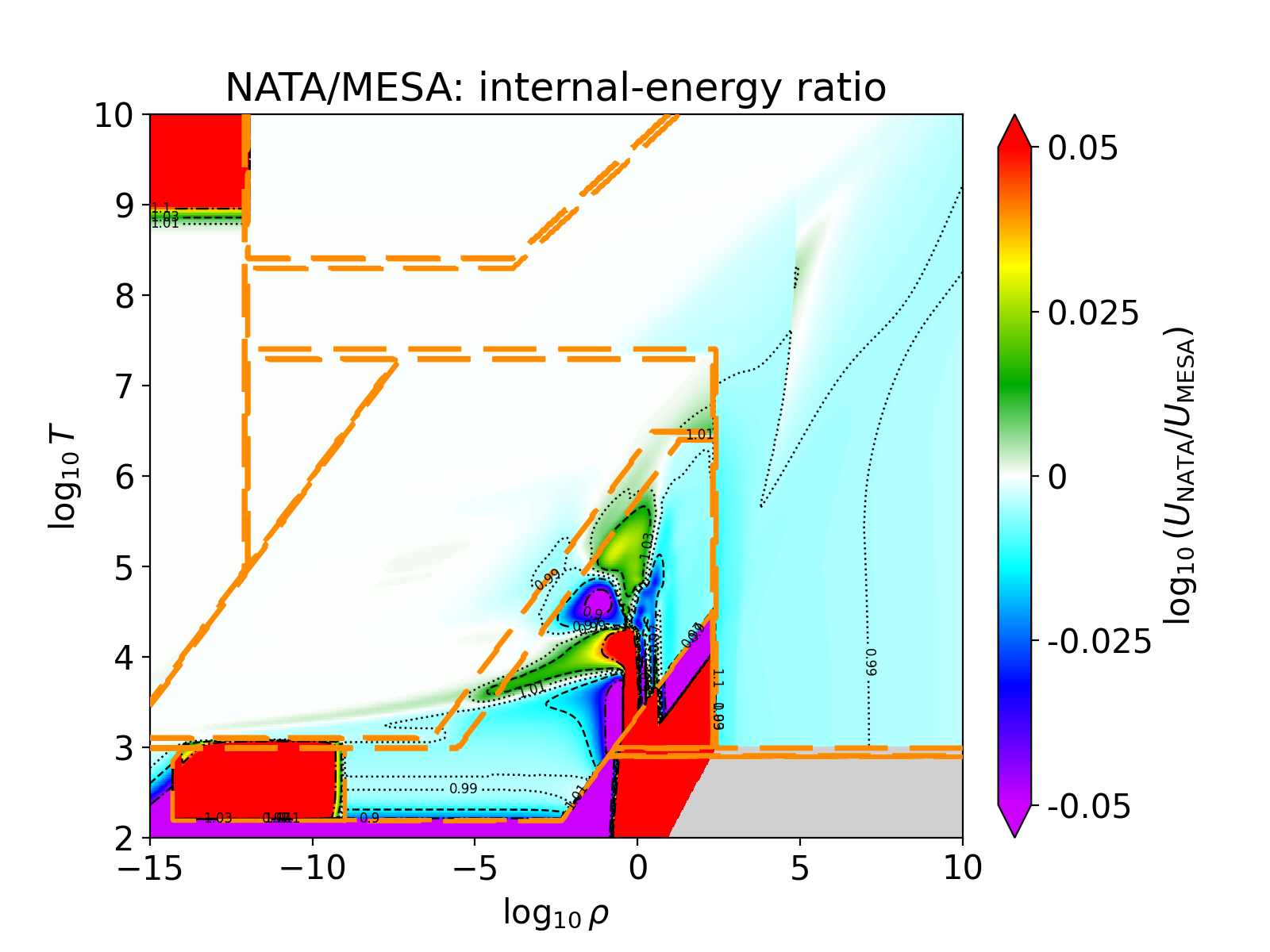}{0.49\textwidth}{(b)}}
  \gridline{\fig{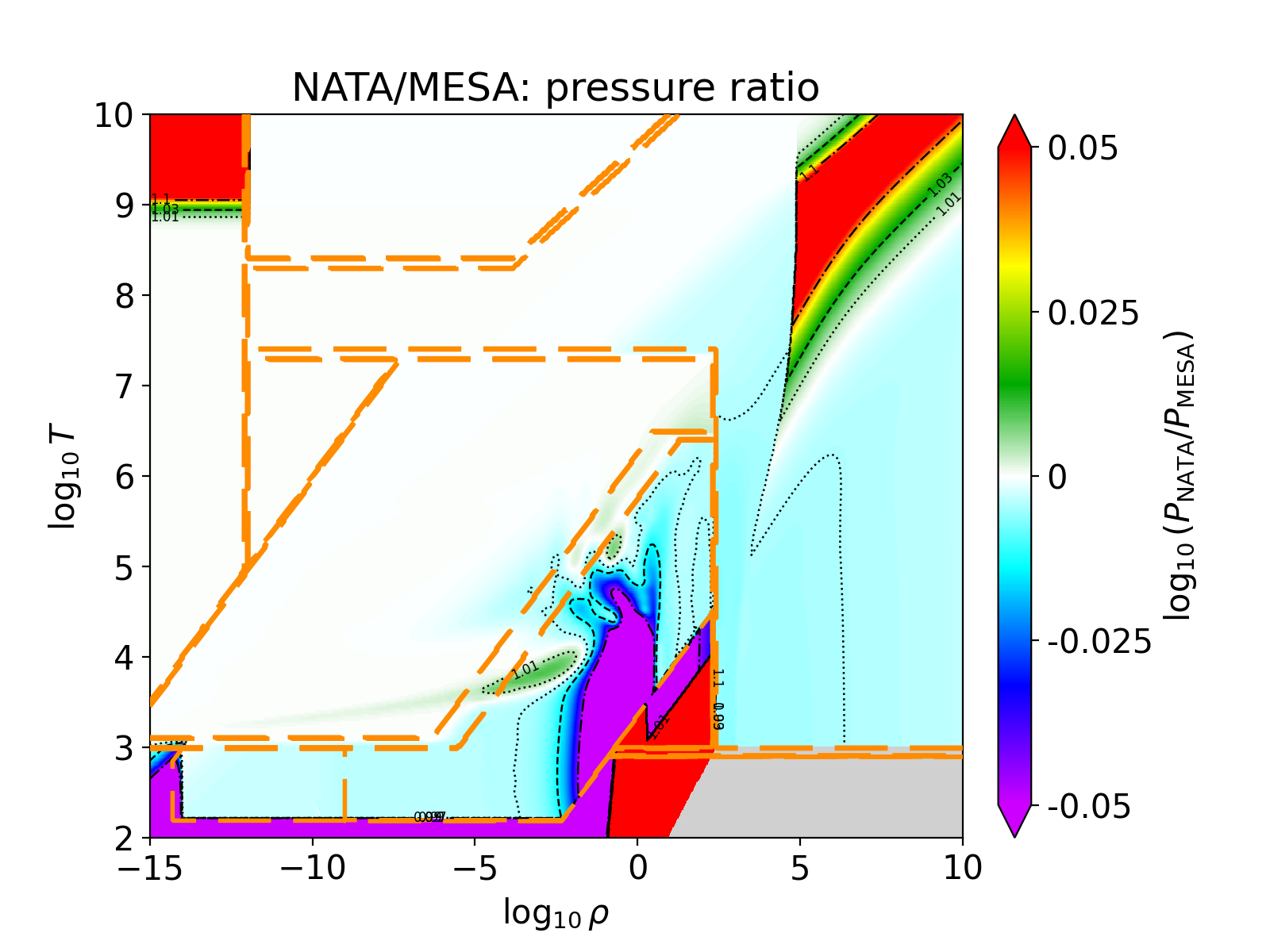}{0.49\textwidth}{(c)}
            \fig{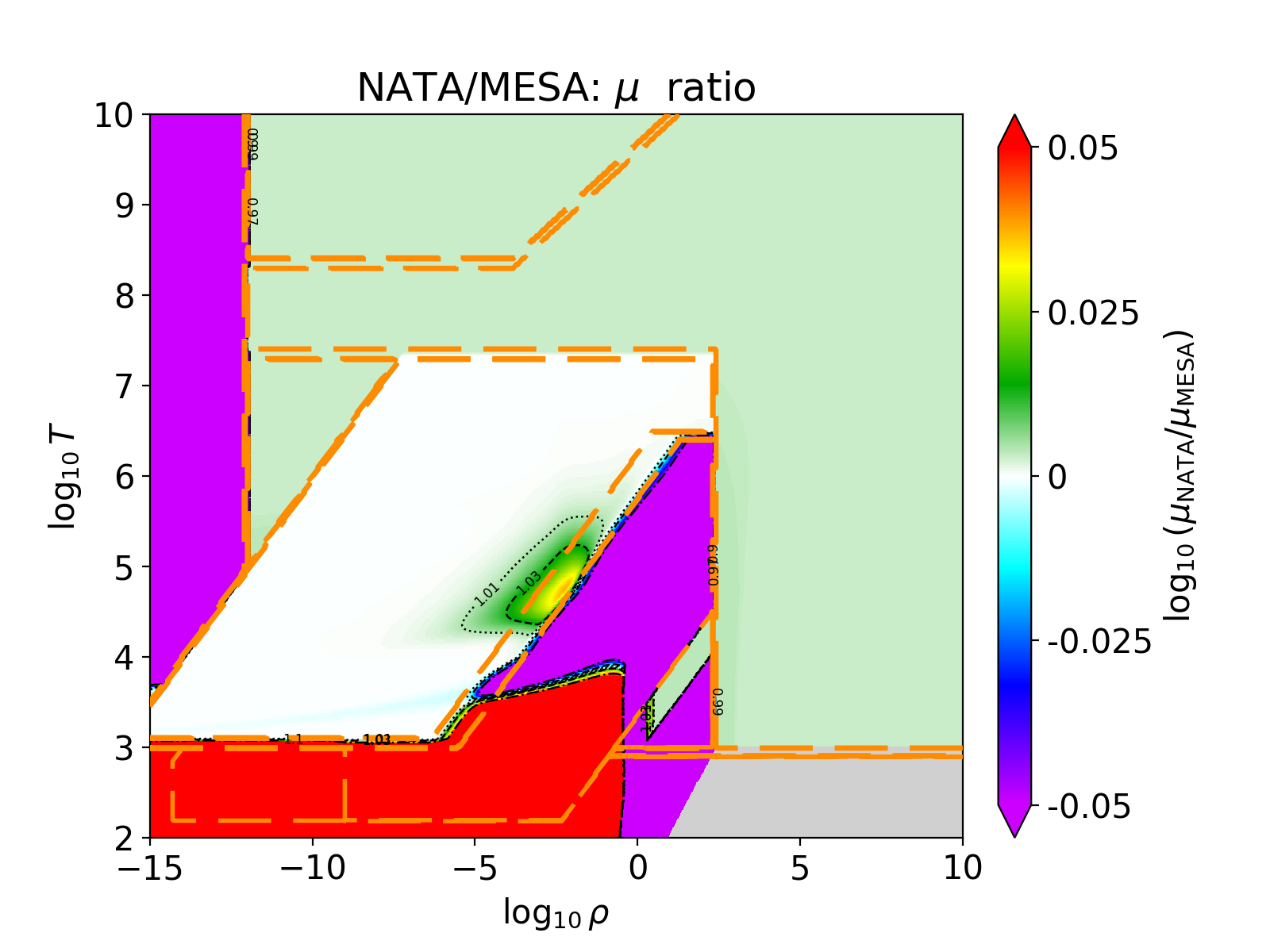}{0.49\textwidth}{(d)}}
  \caption{NATA to MESA comparison for the stellar mixture $X=0.70$, $Y=0.28$,
    $Z=0.02$, in the $\log_{10}(\rho/\mathrm{g\,cm^{-3}})$--$\log_{10}(T/\mathrm{K})$
    plane; all four panels share the same axes.  Panel~(a) is the MESA r26041 EOS
component selection map: the filled colour gives the component EOS that MESA
    uses at each point in the default configuration (ideal, HELM, OPAL/SCVH,
    FreeEOS, Skye, or
    \emph{blend}, the last denoting more than one active component), following the
    legend; dark lines mark the component seams.  PC and CMS are disabled in
    that configuration and therefore do not appear.  Panels~(b), (c), and~(d) show
    $\log_{10}$ of the ratio of NATA to MESA for the internal energy $U$, the
    pressure $P$, and the mean molecular weight $\mu$, respectively.  The diverging
    colour scale saturates at $\pm0.05$ in $\log_{10}$ (a factor $10^{\pm0.05}$,
    about $\pm12\%$); values beyond this range are shown in the end colours (red
    high, magenta low), as marked by the colour-bar arrows.  Black contours trace
    the constant ratio: dotted at $0.99$ and $1.01$, dashed at $0.97$ and $1.03$,
    and dash-dotted at $0.90$ and $1.10$, labelled by value.  Orange long-dashed
    curves reproduce the MESA component seams of panel~(a).  Grey marks points with
    no valid comparison (masked or failed table entries from MESA).%
    \label{fig:mesa_ratio_sm}}
\end{figure*}

\begin{figure*}
  \gridline{\fig{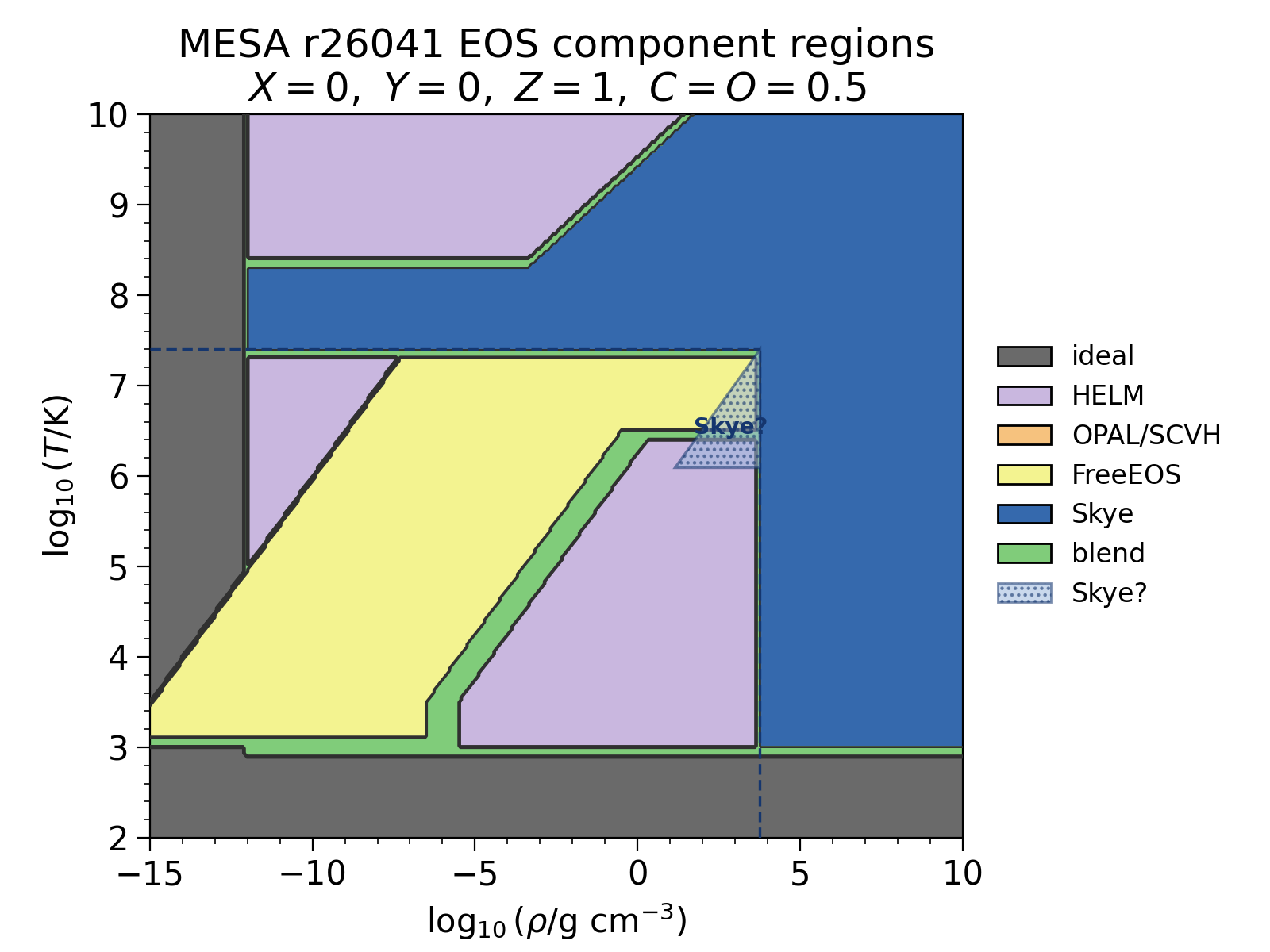}{0.49\textwidth}{(a)}
            \fig{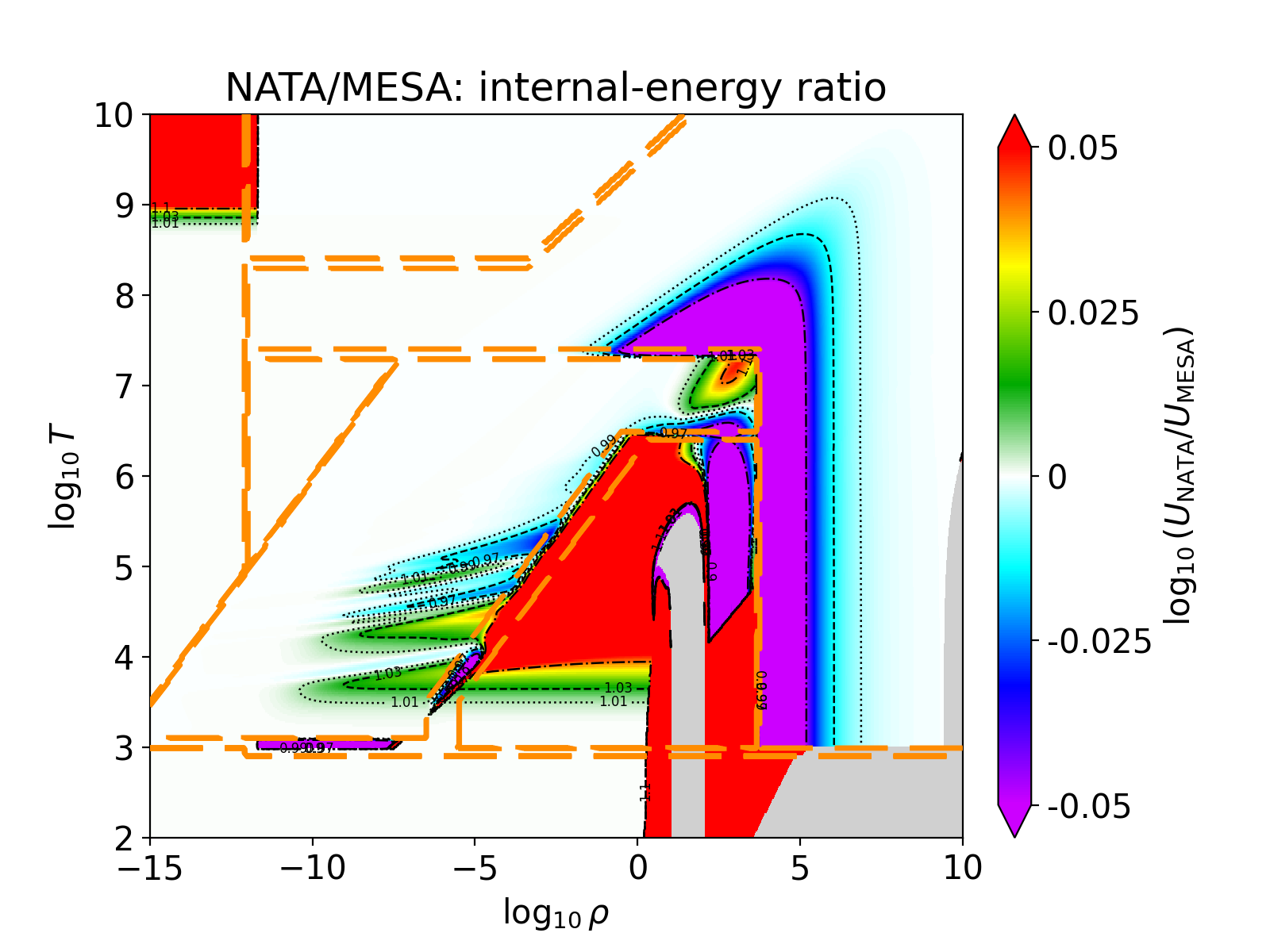}{0.49\textwidth}{(b)}}
  \gridline{\fig{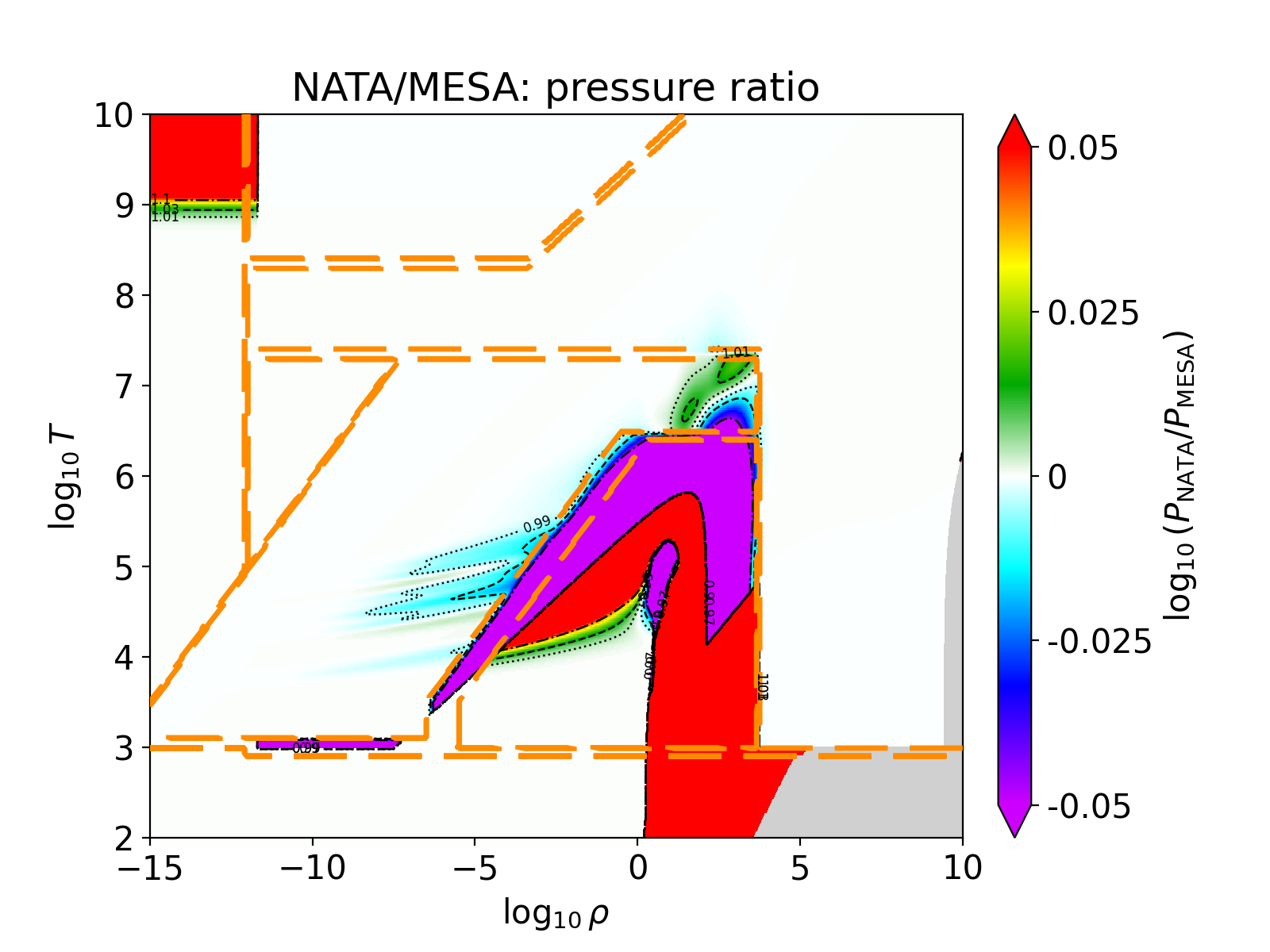}{0.49\textwidth}{(c)}
            \fig{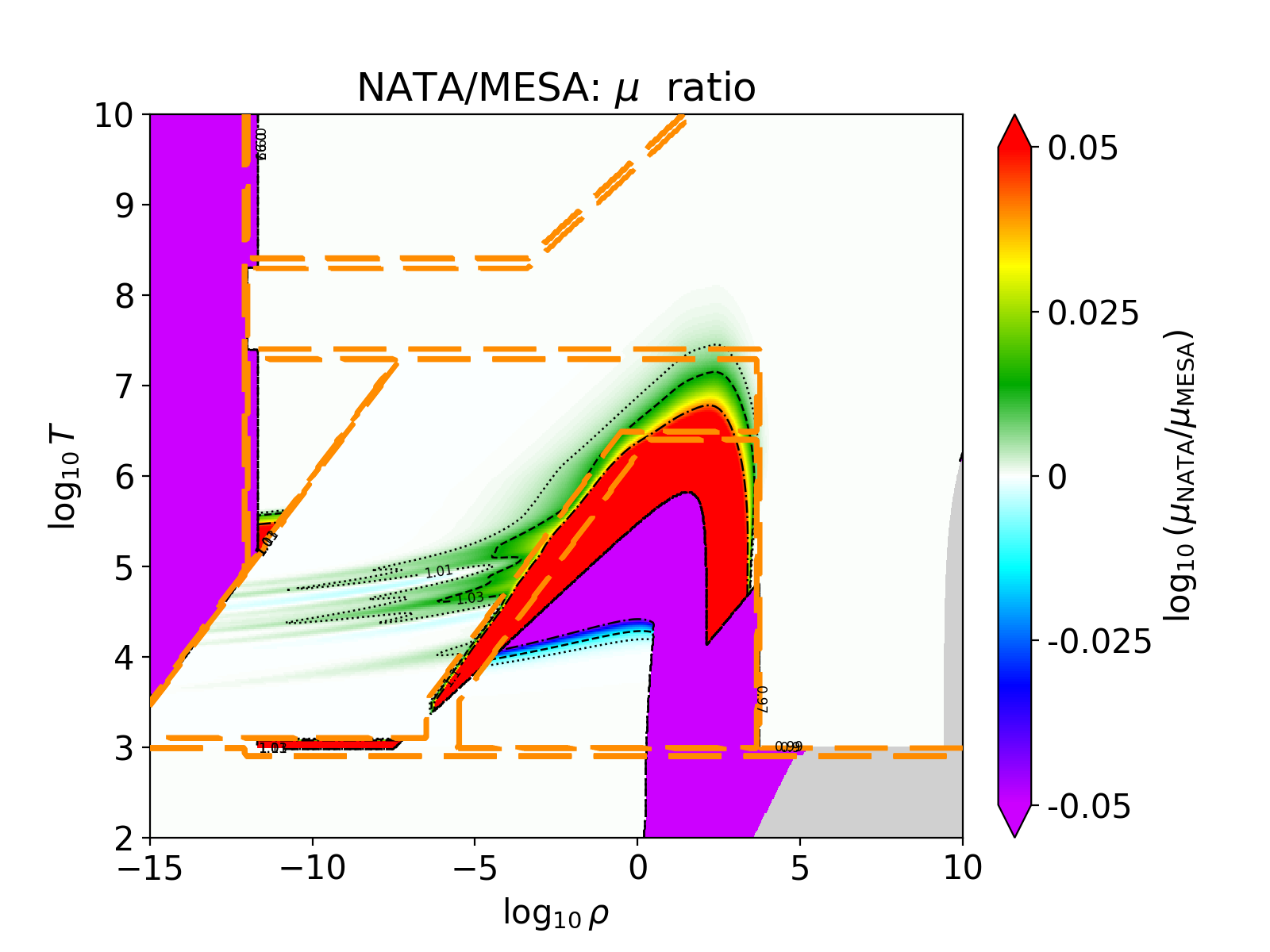}{0.49\textwidth}{(d)}}
  \caption{As Figure~\ref{fig:mesa_ratio_sm}, but for the pure metal composition
    $C=O=0.5$ by mass ($Z=1$, $X=Y=0$).  Panel~(a) is the MESA r26041 EOS component
    selection map; panels~(b), (c), and~(d) show $\log_{10}$ of the ratio of NATA
    to MESA for the internal energy $U$, the pressure $P$, and the mean molecular
    weight $\mu$.  Colour scale, contour levels, the orange component seams, and the
    grey mask are as in Figure~\ref{fig:mesa_ratio_sm}.%
    \label{fig:mesa_ratio_co}}
\end{figure*}

\subsection{Global comparison with the MESA EOS}
\label{sec:mesa_global}

The comparison EOS is the MESA \texttt{eos} module
\citep{Paxton2011,Paxton2013,Paxton2015,Paxton2018,Paxton2019,Jermyn2023}, which
blends several component EOS according to the local thermodynamic conditions:
OPAL \citep{RogersNayfonov2002}, SCVH \citep{Saumon1995}, FreeEOS
\citep{Irwin2012}, HELM \citep{TimmesSwesty2000}, PC \citep{PotekhinChabrier2010},
CMS \citep{ChabrierMazevetSoubiran2019}, and Skye \citep{Jermyn2021}.  We use the default
MESA configuration, in which PC and CMS are disabled and Skye is active, so
neither PC nor CMS contributes to the comparison or to the region maps below.
We map the ratio of the NATA internal energy to the
MESA internal energy across the $(\log\rho,\log T)$ plane.

The blend regions shown on the maps were reconstructed from the MESA
r26041 source.  The mapped regions appear largely the same as those
reported in \citet{Jermyn2023}.  The one addition in our region map is a
part of the SCVH/OPAL region, recently reported on
GitHub\footref{fn:mesa995}, where OPAL/SCVH falls back to HELM without a flag
at densities below which SCVH was not computed.  A second caveat
concerns the Skye boundary.  \citet{Jermyn2021} select Skye where
$\log T > 6.2$ or $\log\rho > 4$, but caution in their Figure~17 that
``the precise shape of the blend between this EOS and the others is more
complicated than a simple cutoff'', which shows a triangular wedge near the corner where the two thresholds meet, in which Skye extends into the region that the r26041 source and \citet{Jermyn2023} label FreeEOS.

In what follows, we first analyse the differences for all components apart from the OPAL/SCVH region, then consider the OPAL/SCVH region separately in \S~\ref{sec:opal_scvh}.

\subsubsection{Stellar mixture.}

\label{sec:mesa_sm}

For the stellar mixture composition, across most of the plane where MESA is reliable the two EOS agree in internal energy to within a few percent; on the fully ionised reliable plane the median ratio difference is $0.05\%$, the ninetieth percentile $0.76\%$, and the maximum $1.1\%$.

The green bands in Figure~\ref{fig:mesa_ratio_sm} are the blend zones, where MESA mixes two adjacent
components through the \texttt{alfa} weighting rather than switching abruptly.
The NATA internal energy is a single continuous expression, so the ratio map shows directly whether MESA stays continuous across each seam.
The transition between HELM and Skye is smooth.  The transition between FreeEOS
and Skye across the constant temperature border carries the same step described
next, but diluted: there the total energy is large and partly radiation, so the
fractional step is only about $0.5\%$ and the border looks nearly smooth.  

At the FreeEOS to Skye border along constant density the step is fully visible:
NATA sits slightly above FreeEOS and slightly below Skye. The step is a particle-count convention: NATA and FreeEOS form number densities from the actual
ion masses $n_i=\rho X_i/m_i$ and return $\mu=0.6210$ and $0.6211$, while Skye
uses the integer convention $m_i=A_i\,u$ and returns $0.61678$, lower by $0.68\%$.
Because ideal pressure scales as $1/\mu$ and ideal thermal energy as $T/\mu$, the
same count difference appears as a $0.68\%$ pressure offset and a
temperature-proportional energy offset wherever Skye is used, with NATA on the
FreeEOS side.

Saturated values, where the difference exceeds a few percent, appear in three
places:
\begin{itemize}
\item the high density, low temperature OPAL region, analysed in more detail in \S~\ref{sec:opal_scvh};

\item regions where the returned MESA component lies outside the physical regime  relevant to the comparison: two HELM regions at $\log T \le 3.1$, and one   further region at low density and high temperature; the low-temperature ideal  EOS also differs systematically, and not all of these regions are documented. In a localised corner the MESA blend falls back to the fully ionised HELM component below $\log T \sim 3$, where the FreeEOS and OPAL/SCVH tables return  HELM values: a fully ionised mean molecular weight ($\mu \sim 0.62$) and  electron fraction in a regime where the physical gas would be neutral or  molecular.
Differences against MESA in this corner therefore arise on the MESA side of the comparison rather than in NATA;

\item a degeneracy wedge in the Skye region, at high density and temperature,  more pronounced in pressure than in internal energy and absent in $\mu$.  At  $\log\rho=5$, $\log T=8$ the pressure ratio is $1.142$ while the energy ratio  is $1.004$ and the $\mu$ ratio is the flat $1.007$ offset, so the wedge is a  pressure feature with almost no energy partner.  
NATA's own $u$, $P$, $S$, and $\mu$ are smooth and monotonic across this region, so the feature appears on the MESA/Skye side of the comparison rather than in NATA.  We discuss it next.

\end{itemize}

The degeneracy wedge sits in the partially degenerate, mildly relativistic
electron regime, $T/T_F \sim 0.02$ to $0.25$ and $x \sim 0.4$ to $1.4$, where
$T_F$ is the electron Fermi temperature and $x=p_F/m_ec$ the relativity parameter
of \S\ref{sec:ref_independent}.
  That it is strong in pressure and weak in internal energy and mean molecular weight locates it in the electron and Coulomb thermodynamics rather than the ionisation state. Across this band, $\eta_e \sim 2$ to $23$,  NATA reproduces the exact ideal finite temperature Fermi gas pressure, with ideal ions and radiation:
against an independent Fermi-Dirac evaluation NATA holds $0.977$ to $0.984$ of the ideal pressure smoothly, a residual of $1.6$ to $2.3\%$ that is the smooth electron exchange and ion Coulomb correction.  Over the same band MESA holds $0.99$ of the ideal pressure up to $\log\rho \simeq 4.8$ and then drops to $0.86$ at $\log\rho = 5.0$, 
so the departure from the analytic electron gas is on the Skye side, not NATA's.
 On the strongly degenerate flank, $T/T_F \lesssim 0.02$, NATA matches the
analytic Chandrasekhar limit to about $2\%$, and there NATA and Skye agree to about a percent ($P_{\rm MESA}/P_{\rm NATA}=0.99$ to $1.01$ at $\log\rho=6$ to $7$), so the wedge fades (Section~\ref{sec:ref_independent}).

Within the degeneracy wedge the departure is thus smooth on the NATA side and appears entirely on the MESA/Skye side.  A sharper feature near $\log\rho \sim 4.8$ appears as a one-cell pressure feature in the comparison: at fixed temperature the ratio $P_{\rm NATA}/P_{\rm Skye}$ rises from $0.99$ to $1.17$ at $\log T=8$ across a single grid cell of $0.025$ in $\log\rho$, between $\log\rho=4.80$ and $4.825$, deepening with temperature, while NATA and the analytic Fermi gas pass through smoothly.  In the comparison it falls where MESA returns Skye and runs close to constant density,
$d\log\rho/d\log T \simeq 0.17$, tracking neither the Coulomb coupling nor the degeneracy crossover.  It is therefore consistent with a table or interpolation boundary on the MESA/Skye side rather than a smooth thermodynamic transition, though we do not resolve its origin within Skye from this comparison.

\subsubsection{Pure metal mixture.}
For the pure metal composition, $C = O = 0.5$, shown in Figure~\ref{fig:mesa_ratio_co}, the agreement is generally very
good with FreeEOS, with high temperature HELM, and with low density Skye. 
The few percent wing shaped depressions relative to FreeEOS, which trace the carbon and oxygen ionisation fronts, are analysed at the end of this subsection.

Across the C/O region the pressure agrees with Skye to better than $0.05\%$
over $95$ per cent of the Skye fully ionised plane,
while the internal energy differs by an additive constant wherever Skye is used
and the gas is fully ionised:
$u_{\rm Skye}-u_{\rm NATA}=1.31\times10^{15}\,{\rm erg\,g^{-1}}$,
constant to a median $0.01\%$ and a ninety-ninth percentile $0.4\%$ of the local energy across the Skye fully ionised plane, excluding a one-cell pressure ridge at $\log\rho\simeq3.76$ of the same character as the feature discussed in \S\ref{sec:mesa_sm}.
Because this constant carries no pressure, and because the density and temperature derivatives of $u$ agree with Skye to better than
$0.5\%$, it is a zero-point offset in Skye's fully ionised energy reference rather
than a thermodynamic error.  It is also not the neutral-atom binding offset: it is
$12.7$ times the C/O ionisation energy.  At the FreeEOS--Skye boundary NATA again
matches FreeEOS, while the offset appears as a step in $u$ at fixed $\rho$ across
the seam, with continuous pressure; for example, at $\log\rho=0$, $u$ jumps by
$1.3\times10^{15}\,{\rm erg\,g^{-1}}$ between $\log T=7.30$ and $7.40$, while
$\log P$ is unchanged to four figures.  This is a hydrodynamics hazard, not merely
a plotting artefact: a code crossing the seam at fixed $\rho$ sees a spurious jump
in $u$ without a pressure jump, and therefore a spurious temperature jump on
inversion.  The offset appears density-dependent only in the energy ratio because
it is divided by the total energy, which is radiation dominated at low density and
electron-degeneracy dominated at high density; the band is deepest at
$\log\rho\simeq2$--$4$, where $u$ is smallest relative to the fixed offset, and
fades on either side.  The green-to-red excess relative to FreeEOS in the
``Skye?'' wedge coincides with the corner where MESA's blend admits a Skye
contribution inside the nominal FreeEOS region (Figure~17 of \citealp{Jermyn2021}),
so part of that excess is the Skye reference offset entering through the blend
rather than a genuine FreeEOS disagreement.

The mean molecular weight behaves as analysed for the stellar mixture
(Section~\ref{sec:mesa_sm}), with one simplification: the particle count
convention that produced the small systematic offset there is negligible here,
below $0.02\%$ against $0.68\%$ for the hydrogen rich mixture, because the
$^{12}$C and $^{16}$O masses are essentially integer, so $\mu$ agrees with the
reliable MESA components in the fully ionised regions.

Two regions lie outside the reference domains.  The high density, $\log T \le 6$ region that
saturates in the $\mu$ ratio (purple, NATA is below MESA) lies outside the MESA
coverage documented by \citet{Jermyn2023} (their Figure~7), so the ratio there is
meaningless.  The low temperature HELM region is likewise outside the regime of the comparison; it shows as
the saturated red region in the internal energy ratio at $\log\rho$ from about
$-5$ to $5$ and $\log T \le 6$.

These wings are the NATA/MESA internal energy ratio departing from unity along
the carbon and oxygen first ionisation fronts.  At $\log\rho=-5$ the departure has
two parts (Table~\ref{tab:co_wings}):  a narrow feature near $\log T \simeq 3.9$ to
$4.1$, where the mean molecular weights differ by up to a factor of six and the
ratio falls to $0.81$, and a broader oscillation of a few percent between
$\log T \simeq 4.2$ and $5.4$, reaching $1.08$ near $\log T \simeq 4.4$ and $0.96$
near $\log T \simeq 4.7$.  Above $\log T = 5.6$ the two agree to better than
$5\times10^{-4}$.
At these densities electron degeneracy and packing effects are
negligible, so the EOS should approach the ideal Saha ionisation balance for the
adopted partition functions and ionisation potentials.  We evaluate the carbon and
oxygen balance independently of NATA from the Saha equation \citep[e.g.][their
Eq.~5-16]{Mihalas1978}, see also \citet[][\S104]{LL5},
\[
\frac{n_{i+1}\,n_e}{n_i}
 = \frac{2\,U_{i+1}(T)}{U_i(T)}
   \left(\frac{2\pi m_e k_{\rm B} T}{h^2}\right)^{3/2}
   e^{-\chi_i / k_{\rm B}T},
\]
where $n_i$ is the number density of ionisation stage $i$, $U_i(T)$ its internal
partition function, $\chi_i$ the ionisation potential of the $i\rightarrow i+1$
transition, and the factor $2$ the electron spin statistical weight.  We solve this
coupled set for all carbon and oxygen stages simultaneously with charge neutrality
$n_e = \sum_i i\,(n_{{\rm C},i} + n_{{\rm O},i})$, using the same potentials
$\chi_i$ and partition functions $U_i(T)$ that NATA carries and no continuum
lowering, and define the free electron fraction $Y_e = n_e\,m_{\rm u}/\rho$.  This
test therefore verifies that NATA solves its own ideal C/O ionisation ladder
correctly; it does not by itself establish that the adopted $U_i(T)$ are identical
to those of the reference.

The three $Y_e$ are obtained as follows.  For NATA it is the free electron density
of the converged charge neutrality root.  For the independent Saha balance it is
the solution of the equation above.  For the reference (MESA) it is recovered from the
tabulated mean molecular weight, $Y_e = 1/\mu - (X_{\rm C}/A_{\rm C} +
X_{\rm O}/A_{\rm O})$, since $\mu = \rho / [\,m_{\rm u}(n_{\rm ion}+n_e)\,]$ with
$n_{\rm ion}$ fixed by composition; the fully ionised limit $\mu \rightarrow 1.745$
confirms the convention.

\begin{table}[h]
\centering
\begin{tabular}{lcccc}
\toprule
$\log T$ & $Y_{e,\rm NATA}$ & $Y_{e,\rm Saha}$ & $Y_{e,\rm MESA}$ &
 $U_{\rm NATA}/U_{\rm MESA}$ \\
\midrule
$4.0$ & $0.006232$ & $0.006101$ & $0.06676$ & $0.8519$ \\
$4.2$ & $0.048416$ & $0.047737$ & $0.05000$ & $1.0282$ \\
$4.5$ & $0.101233$ & $0.100401$ & $0.10109$ & $1.0635$ \\
$4.7$ & $0.164985$ & $0.164091$ & $0.17448$ & $0.9561$ \\
$5.0$ & $0.284210$ & $0.283663$ & $0.28766$ & $1.0010$ \\
$5.2$ & $0.338498$ & $0.338190$ & $0.34634$ & $0.9687$ \\
$5.4$ & $0.352965$ & $0.352938$ & $0.35419$ & $0.9986$ \\
\bottomrule
\end{tabular}
\caption{Free electron fraction and internal energy ratio at $\log\rho = -5$ for
the $X_{\rm C} = X_{\rm O} = 0.5$ mixture.  NATA tracks the independent Saha
balance to within $2.2$ percent at every temperature, the residual being the
  Stewart-Pyatt depression that the independent balance omits by construction.
The MESA table departs from} the Saha value through the wings, by a few percent near $\log T \simeq 4.7$
and by an order of magnitude at $\log T = 4.0$, and the $U_{\rm NATA}/U_{\rm ref}$
departure follows the difference in free electrons.
\label{tab:co_wings}
\end{table}

Along $\log\rho = -5$, NATA reproduces the independent Saha $Y_e$ to within
$2.2$ percent at every temperature and to better than one percent at and above
$\log T = 4.5$ (Table~\ref{tab:co_wings}).  The independent balance uses the
same completed partition functions and ionisation potentials as the production
solve, and omits only the ground-state occupation dissolution, the continuum
lowering, and the electron-degeneracy correction to the Saha ratios.  With the
Stewart-Pyatt depression switched off the two agree to a few parts in $10^{5}$, so
the residual is the depression alone: it raises the free electron fraction by
$2.15$ percent at $\log T = 4.0$ and by $0.01$ percent at $\log T = 5.4$ as the gas
approaches full ionisation.
The MESA table instead departs from the Saha value through the wings, by a
few percent at $\log T \simeq 4.6$ to $5.2$ and by an order of magnitude at
$\log T = 4.0$, where it returns $Y_e = 0.067$ against the Saha $0.006$.
This suggests that the MESA table, which uses its FreeEOS component here, is not
following the ideal C/O Saha limit in this region.
The energy ratio departs from unity in both directions.  Across the
oscillating band it falls below unity where the mean molecular weight ratio
$\mu_{\rm NATA}/\mu_{\rm MESA}$ rises above it, the two signed departures
correlating at $-0.77$.  At fixed composition the difference in $\mu$ reflects
the difference in free-electron abundance, so the two EOS differ in energy where
they differ in ionisation.  The wings are therefore not evidence that NATA
misplaces the C/O ionisation.

\subsection{The OPAL/SCVH region}
\label{sec:opal_scvh}

\begin{figure*}
\centering
\includegraphics[width=0.49\textwidth]{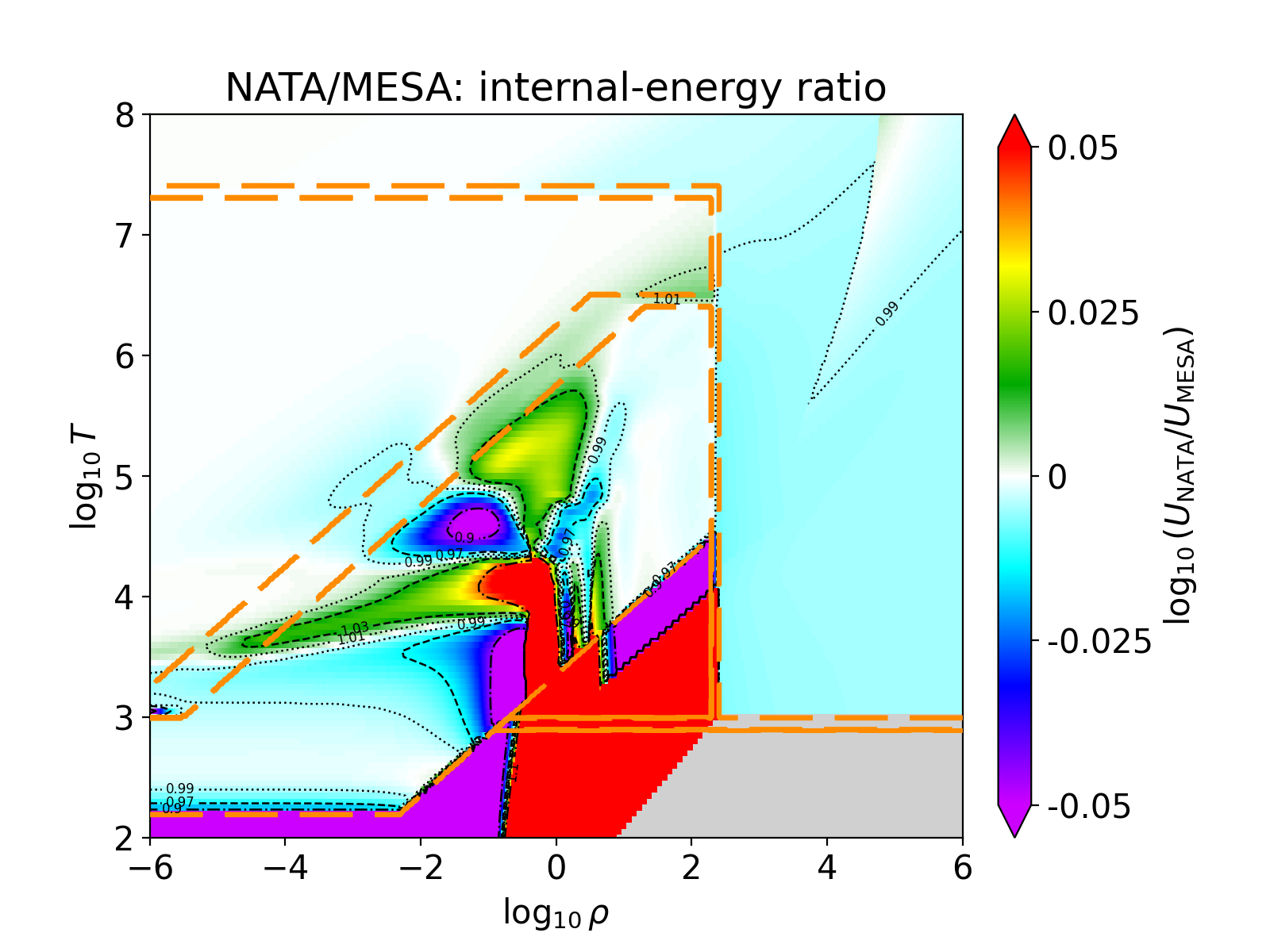}
\includegraphics[width=0.49\textwidth]{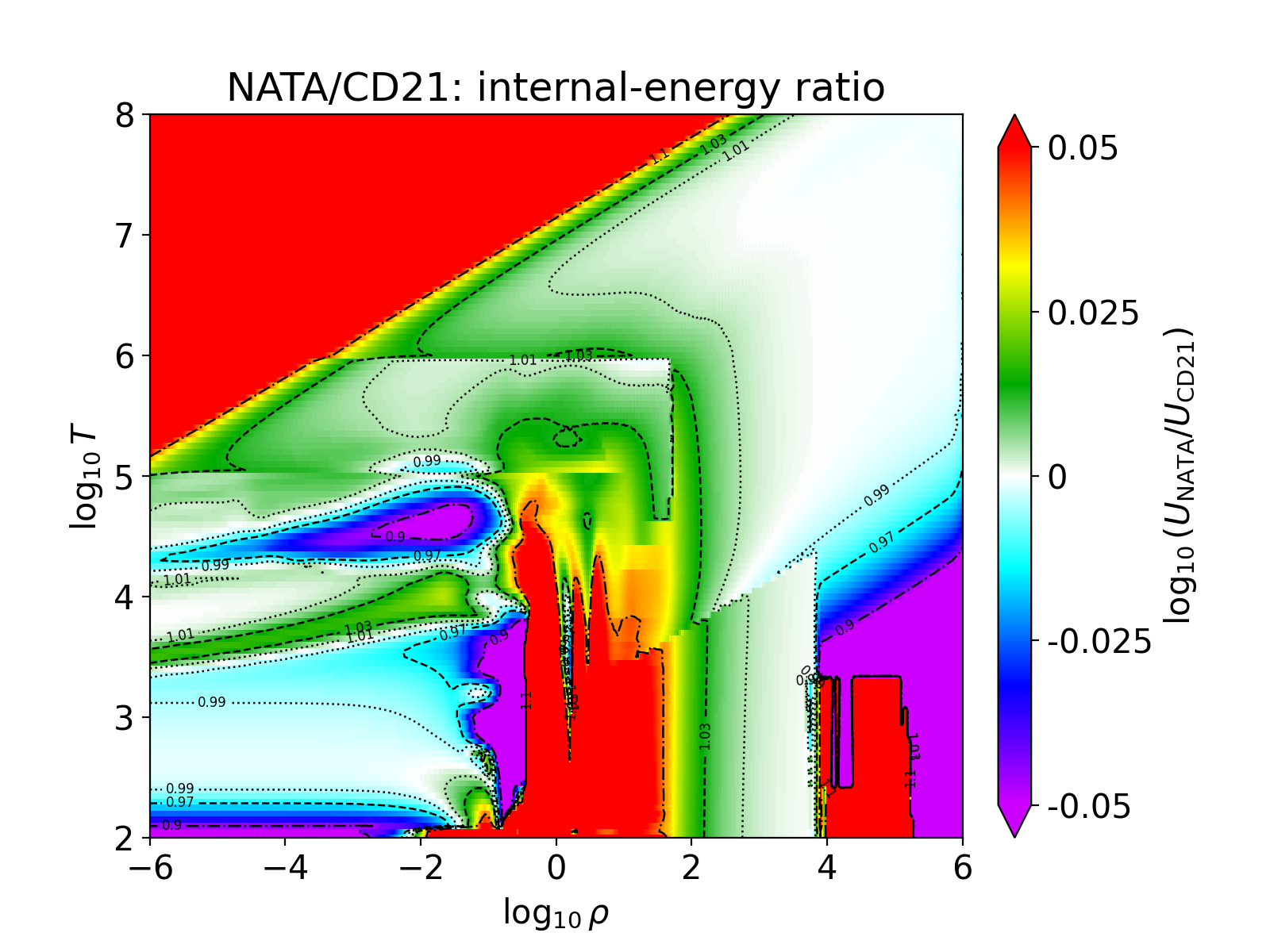}
\caption{Internal energy comparison in the OPAL/SCVH region, for the hydrogen
helium mixture $Y=0.275$, $Z=0$.  Both panels show
$\log_{10}(U_{\rm NATA}/U_{\rm ref})$ over the $(\log\rho,\log T)$ plane on a
fixed colour scale from $-0.05$ to $0.05$ (green where NATA is higher, blue to
purple where it is lower, saturating to red and purple beyond $\pm0.05$), with
dashed and dotted contours of the linear ratio at $0.9$, $0.97$, $0.99$, $1.01$,
$1.03$, and $1.1$.  Left: NATA against MESA's blend, which here is OPAL/SCVH with
the FreeEOS corner that falls in the same region; the orange dashed lines mark
the MESA component boundaries and the grey region at high density and low
temperature is outside the MESA table.  Right: NATA against the CD21 hydrogen
helium tables \citep{ChabrierDebras2021}.  
The features discussed in the text are the
$\mathrm{H_2}$ dissociation ridge at low density, the partial and pressure
ionisation band with its paired deficit and excess lobes near the Mott
transition, and, in the right panel,
the regions of the CD21 table that its authors flag as unphysical, the low density edge and the cold dense corner, and the interpolation structure noted in \S\,\ref{sec:opal_scvh}.}
\label{fig:ratio_HHe}
\end{figure*}

In this region the comparison is between equations of state built on different
and equally legitimate foundations, and it is worth stating the distinction
plainly at the outset.  NATA, like FreeEOS \citep{Irwin2012}, is a chemical
picture equation of state: the bound species ($\mathrm{H}$, $\mathrm{H^+}$,
$\mathrm{He}$, $\mathrm{He^+}$, $\mathrm{He^{++}}$, $\mathrm{H_2}$, and the minor
molecular ions) are carried explicitly, in Saha and dissociation equilibrium with
measured atomic and molecular constants and an occupation probability continuum
lowering for pressure ionisation.  OPAL \citep{Rogers1996,RogersNayfonov2002}
works in the physical picture, where no bound species is defined in advance and
ionisation emerges from an activity expansion of the interacting electron ion
system; SCVH \citep{Saumon1995} is a dedicated low temperature hydrogen helium
equation of state; and CD21 \citep{ChabrierDebras2021} builds on the ab initio
free energies of CMS19 \citep{ChabrierMazevetSoubiran2019} with an additive
hydrogen helium mixing law, adding the non-ideal interaction terms from the
quantum molecular dynamics simulations of \citet{MilitzerHubbard2013}.

These references carry physics that NATA's ideal chemical mixture does not attempt
to reproduce.  The physical picture captures pressure ionisation and the
electron ion non-ideality directly, without committing in advance to a list of
bound states; the ab initio tables carry genuine non-ideal hydrogen helium
interactions in the dense and degenerate regime; and the dedicated low temperature
treatments resolve the molecular and condensed phase behaviour in detail.  NATA is deliberately lighter there, and we make no claim otherwise.  We therefore do not expect, and do not aim for, agreement with OPAL or CD21 wherever the difference is the formalism itself: forcing it would require importing the physical picture or ab initio non-ideal treatment and would defeat the purpose of a transparent chemical picture equation of state.  What can be done, and what we do here, is to decode the residual features one by one, separating those that are the picture  choice, and so are expected and irreducible within the chemical picture, from any  that would instead point to a correctable datum on the NATA side.  In every case examined below the verifiable standard is the textbook ideal limit, which both pictures must share where the gas is dilute, and it is against that limit, not against either reference, that NATA is tested.

Here we compare a hydrogen helium mixture at $Y=0.275$, $Z=0$ on the
$(\log T,\log\rho)$ plane against two references, MESA's OPAL/SCVH (with the FreeEOS corner in the same region) and the \citet{ChabrierDebras2021} hydrogen helium equation of state (CD21); see Figure~\ref{fig:ratio_HHe}.  The two references differ substantially from each other, not only from NATA, so the features that persist against both are the ones we discuss.

CD21 extends the additive volume law (ideal mixing) tables of CMS19 by adding the non-ideal hydrogen helium interaction terms from \citet{MilitzerHubbard2013} (MH13).  Its table spans $\log T\in[2,8]$ and $\log\rho\in[-6,6]$ at a uniform $0.05$ dex step, with pressure, internal energy, and entropy in $\mathrm{GPa}$,
$\mathrm{MJ\,kg^{-1}}$, and $\mathrm{MJ\,kg^{-1}\,K^{-1}}$, which we convert to cgs with the common factor $10^{10}$ before forming ratios; $T$ and $\rho$ are already in K and $\mathrm{g\,cm^{-3}}$.

Several features of the residual maps below trace to documented properties of the CD21 table rather than to NATA.  First, the table is rectangular and not everywhere physical: 
\citet{ChabrierMazevetSoubiran2019} caution that entries at the lowest densities, and values in the solid phase or beyond the first order quantum ion correction, are unphysical (their Figs.~1 and 16), so residuals along the low density edge and in the cold dense corner reflect the reference fill and we exclude them, since the CD21 tables we use inherit these flagged regions from the additive volume law tables they extend;
any comparison at $\log\rho<-6$ lies outside the tabulated domain and is not made.
Second, the table is spline interpolated, and CD21 state that numerical oscillations and interpolation flaws are to be expected, so fine scale oscillatory structure in the ratio maps that matches no ionisation, dissociation, or degeneracy feature of NATA is reference side interpolation noise; in the stellar regime CD21 themselves note the non-ideal contributions are inconsequential and recommend the ideal mixing CMS19 tables to avoid it, and there NATA, an ideal chemical picture mixture, should track CMS19/CD21 to within
that noise.  Third, in the brown dwarf domain at low temperature and high density a genuine physical residual is expected, since the non-ideal interaction term CD21 carries is absent in NATA; this regime also has the largest reference uncertainty, as the MH13 simulations cover a limited temperature density domain.

\subsubsection{The mean molecular weight.}
In the OPAL/SCVH region $\mu$ differs markedly from NATA, by as much as several
tens of percent and with both signs across the field.  Part of this is the
genuine chemical versus physical picture difference in the ionisation state
through the pressure ionisation region, where the two formalisms place the
ionisation fronts at marginally different density and width.  The size and
pattern of the difference suggest, however, that the reference may also count particles differently.  The CD21 tables do not tabulate $\mu$, and the OPAL tables likely do not either, so MESA recovers $\mu$ from the total pressure; the convention in that recovery need not match ours,
$\mu=\rho/[\,m_u(n_{\rm ion}+n_e)\,]$.  The behaviour is essentially the same as for the stellar mixture, so we do not produce a separate figure here and refer to the $\mu$ panel of Figure~\ref{fig:mesa_ratio_sm}.

\subsubsection{The dissociation front.}

A positive green ridge runs along the $\mathrm{H_2}\leftrightarrow 2\,\mathrm{H}$ dissociation front from $\log T \approx 3.55$, $\log\rho = -6$ toward higher
density at slightly rising temperature, where the internal energy is governed by
the dissociation energy $D_0$ released as the bond breaks; the gas is neutral
here, so no ionisation potential enters.  The balance is the diatomic law of mass
action, in which the $\mathrm{H_2}$ internal partition function divides the
equilibrium constant, $K=n_{\mathrm H}^2/n_{\mathrm{H_2}}\propto
Q_{\mathrm{H_2}}^{-1}\,e^{-D_0/k_{\rm B}T}$, so the level density near dissociation
controls the molecular fraction and hence the energy stored in dissociated atoms.
Against FreeEOS, which carries a spectroscopic $\mathrm{H_2}$ partition function,
NATA reproduces the front to high precision and the ridge nearly vanishes
(Fig.~\ref{fig:mesa_ratio_sm}, or the very left on the left plot in Fig.~\ref{fig:ratio_HHe}); the two chemical picture codes share the same molecular physics and the measured dissociation energy and agree along the front to within the interpolation noise,
The residual positive ridge is therefore seen only against the ab initio tables.

At the low-density end of the front the test is clean, because non-ideal effects
are negligible.  At $\log T = 3.55$ ($T = 3548$~K) and $\log\rho = -6$ the gas is
neutral, so there is no Coulomb term, and the only relevant non-ideal
contribution is the excluded volume of the neutral particles.  If the hydrogen is
molecular, the $\mathrm{H_2}$ packing fraction is
\[
\eta = \frac{\pi}{6}\,n_{\mathrm{H_2}}\,\sigma_{\mathrm{H_2}}^3
\simeq {\rm few}\times10^{-6},
\]
using $\sigma_{\mathrm{H_2}}\simeq 2.9$~\AA\ for the hard-sphere diameter
\citep{HirschfelderCurtissBird1954}.  The leading correction to the ideal
pressure, the second virial term $B_2 n = 4\eta$ for hard spheres
\citep[][\S74]{LL5}, is therefore of order $10^{-5}$, some five orders of
magnitude below the ideal term.  Any correct equation of state must therefore
reduce here to the textbook ideal dissociation equilibrium, to this accuracy.

We evaluate that balance independently of NATA from the diatomic law of mass
action \citep[][\S102]{LL5}, with the internal partition function summed directly
over the bound rovibrational levels in the manner of \citet{BarklemCollet2016}.
We use the $X\,^1\Sigma_g^+$ constants of \citet{HuberHerzberg}
($\omega_e = 4401.213$, $\omega_e x_e = 121.336$, $\omega_e y_e = 0.8129$,
$B_e = 60.853$, $\alpha_e = 3.0622\,\mathrm{cm^{-1}}$) and the spectroscopic
dissociation energy $D_0 = 36118\,\mathrm{cm^{-1}} = 4.478$~eV, consistent with
the precise value $36118.0696\,\mathrm{cm^{-1}}$ of \citet{Liu2009}.
At $3548$~K the direct sum
gives $Z^{\rm int}_{\mathrm{H_2}} = 26.2$, about five percent above the rigid
rotor plus harmonic oscillator value $25.0$.  NATA carries the corresponding
direct rovibrational sum, giving the same partition function to better than
$0.1\%$, so the partition function is not the source of any residual here.
Solving the mass action law for the molecular fraction and adding helium as a
neutral spectator for the pressure gives $f_{\mathrm{H_2}} =
2n_{\mathrm{H_2}}/n_{\mathrm{H,tot}} = 0.369$ and
$P = 1.934\times10^{5}\,\mathrm{dyn\,cm^{-2}}$.

\begin{table}[h]
\centering
\begin{tabular}{lccc}
\toprule
 & $f_{\mathrm{H_2}}$ & $P\ [\mathrm{dyn\,cm^{-2}}]$ & $U\ [\mathrm{erg\,g^{-1}}]$ \\
\midrule
Independent ideal & $0.369$ & $1.933\times10^{5}$ & -- \\
NATA              & $0.369$ & $1.933\times10^{5}$ & $1.326\times10^{12}$ \\
CMS19/CD21        & $0.402$ & $1.898\times10^{5}$ & $1.280\times10^{12}$ \\
\bottomrule
\end{tabular}
\caption{Ideal dissociation balance at $\log T = 3.55$, $\log\rho = -6$,
$X = 0.725$, $Y = 0.275$, $Z = 0$.  The independent row is the ideal mass-action
solve; the NATA and CD21 rows are read from their tables, with $f_{\mathrm{H_2}}$
recovered from the tabulated $\mu$ for NATA and from the tabulated pressure for
CD21, since neither table lists the molecular fraction directly and the gas here
is ideal and neutral.  NATA reproduces the independent textbook result; CMS19/CD21
retains more $\mathrm{H_2}$.}
\label{tab:dilute_ridge}
\end{table}

NATA reproduces the independent calculation to better than $0.1\%$ in both
$f_{\mathrm{H_2}}$ and pressure (Table~\ref{tab:dilute_ridge}), while CMS19/CD21
retains about $10\%$ more $\mathrm{H_2}$, hence about $2\%$ fewer free particles
and the lower pressure.  Reproducing the CD21 molecular fraction from the same
mass action law would require an effective dissociation energy
$D_0^{\rm eff} \simeq 4.54$~eV, that is $0.06$~eV above the measured value, which
no standard ideal gas term produces.  The excess atomic hydrogen in NATA carries
the dissociation energy, so the particle excess maps directly onto
$U_{\rm NATA}/U_{\rm CD21} = 1.326/1.280 = 1.036$, the ridge seen on the map.  The
same comparison made between FreeEOS and the reference tables, with NATA absent,
reproduces this ridge against CD21 and not against the physical picture OPAL,
confirming that the residual is a property of the ab initio table rather than of
NATA: it reflects the dilute edge convention of that table, which CD21 themselves
caution against using in this regime in favour of the ideal mixing CMS19 tables.
In this dilute limit the correct standard is the chemical picture with the
measured dissociation energy, and forcing agreement with CD21 would move NATA off
the spectroscopic $D_0$ and would be a degradation, not a correction.

\subsubsection{The partial and pressure ionisation band.}

The broad residual band at $\log T \simeq 4$--$6$ and
$\log\rho \simeq -3$--$+2$ is the H/He partial-ionisation region, extending on
the dense side into the regime where pressure (Mott) ionisation and electron
degeneracy become important.  In this regime the internal energy is dominated by
the ionisation reservoir and by how rapidly that reservoir is populated across
the front.  The front is placed and broadened differently in each EOS formalism:
NATA treats the bound species explicitly in Saha equilibrium with
occupation-probability continuum lowering, OPAL represents the same physics
through its activity expansion, and CD21 through ab initio free energies with an
additive mixing prescription.  Differencing a chemical-picture energy against a
physical-picture or ab initio energy therefore does not produce a single offset.
Instead, small displacements in the ionisation front are superposed on steep,
partly compensating ionisation, excitation, and Coulomb terms, producing paired
deficit and excess lobes.

Over most of the band the residual is at the percent level: against the MESA OPAL
map the median is $1.0\%$ and the ninetieth percentile is $7.0\%$.  The largest
structure occurs in the paired lobes around the Mott transition near
$\log\rho\simeq -2$--$0$: NATA is low by up to about $13\%$ on the low-density
side and high on the dense side, with the ratio rising steeply near the foot of
the band at $\log T\simeq4$, where the absolute internal energy is small.  In the
NATA/CD21 comparison the same feature appears as the green ridge through
$\log T\simeq5$--$6$, where NATA is $3$--$6\%$ high, together with the paired blue
deficit and green excess across the Mott transition.  In the NATA/MESA(OPAL) map
the band extends over $\log\rho\simeq -3$--$+2$, with its deficit lobe near
$\log T\simeq4.5$, $\log\rho\simeq -2$, where the ratio falls to about $0.90$.

This morphology is the signature of the EOS picture choice, not of an erroneous
species datum.  It can be reduced by improving the continuum-lowering and
free-energy consistency, but it cannot be made to disappear against all
references without adopting, or tuning to, their particular front placement.  In
the dilute parts of the front where the ideal limit is well defined, NATA agrees
with the corresponding textbook mass-action/Saha balances, so part of the lobe
reflects a reference-side formalism difference rather than a NATA error.  The
dilute ionisation front is checked in the same spirit as the dilute dissociation
test above (Table~\ref{tab:dilute_ridge}), where the ideal limit is reached
directly.  
In separate diagnostic tests the $F_{\rm C}$-consistent continuum-lowering option (\S\ref{sec:nata_eos}) reduced the amplitude of this structure relative to the earlier empirical treatment, but it was not adopted for the reported tables because of the multi-root, positive-feedback behaviour noted there; the production tables use the Stewart-Pyatt depression throughout.

\subsubsection{Molecular regime versus SCVH.}

The molecular part of NATA is designed to capture the molecular contribution to
the bulk thermodynamics rather than to provide a complete low-temperature
atmospheric chemistry network.  The only molecular reservoir that is
energetically important for the hydrogen-rich applications considered here is
neutral $\mathrm{H_2}$, because its dissociation changes both the particle number
and the internal-energy budget in the same cool-envelope regime where hydrogen
recombination becomes important.  We therefore treat $\mathrm{H_2}$
spectroscopically: its internal partition function is evaluated as a direct sum
over the bound rovibrational levels of the $X\,^1\Sigma_g^+$ ground electronic
state, following the construction of \citet{BarklemCollet2016}, referenced to
$v=0$, $J=0$.  This gives a single consistent function for the
$\mathrm{H_2}\leftrightarrow\mathrm{H}+\mathrm{H}$ mass-action balance, the
molecular heat capacity, and the dissociation-energy reservoir, while avoiding
the artificial high-temperature divergence of a rigid-rotor/harmonic-oscillator
partition function.

This treatment is intentionally less extensive than a dedicated low-temperature
hydrogen-helium EOS such as SCVH \citep{Saumon1995}, but the omitted molecular
structure is not a leading-order bulk-EOS term.  The molecular ions
$\mathrm{H_2^+}$, $\mathrm{He_2^+}$, and $\mathrm{HeH^+}$ are retained through
their ground-state dissociation energies and statistical weights rather than full
rovibrational partition functions.  In the NATA solutions they remain trace
species, with $n(\mathrm{H_2^+})/n(\mathrm{H})\lesssim10^{-3}$ even where
$\mathrm{H_2}$ dissociation overlaps the onset of hydrogen ionisation; their
omitted internal excitation therefore gives a negligible correction to $P$, $U$,
and $S$.  Likewise, metal-bearing molecules such as CO, $\mathrm{H_2O}$, and OH
are not included.  At $Z\simeq0.02$ they are important opacity and atmospheric
chemistry species, but they do not provide a bulk thermodynamic reservoir
comparable to $\mathrm{H_2}$ dissociation or H/He ionisation.  This hierarchy is
the one relevant for cool, partially molecular envelopes in stellar mergers and
common envelope binaries, where the dominant EOS-sensitive energy reservoirs are
$\mathrm{H_2}$ dissociation and H/He recombination.

\section{Conclusions}
\label{sec:conclusions}

\begin{deluxetable}{ll}
\tablecaption{Reliability of the NATA EOS by region of the
$(\log\rho,\log T)$ plane.\label{tab:reliability}}
\tablehead{\colhead{Region} & \colhead{Status}}
\startdata
Fully ionised, non-pair-dominated & validated against analytic limits and reference EOSs in domain \\
Dilute molecular H/He & validated against ideal mass action \\
Partial ionisation & residuals expected; validated against the dilute Saha/mass-action limits \\
Cold dense, strongly coupled & continuity and regularisation only \\
Hot rarefied pair wedge & $\mu$ diagnostic only; pairs not coupled to the composition solve \\
\enddata
\tablecomments{``Validated'' means checked against an independent standard:
an analytic limit, an ideal mass-action or Saha balance, or a reference EOS
within its own domain.  ``Regularised'' means the EOS returns continuous values
but they are numerically conditioned rather than independently validated.
Within the partial-ionisation region, the bounded C/O pressure-ionisation
band identified in \S\ref{sec:nata_eos} is a mechanically unstable single-phase
continuation.}
\end{deluxetable}

We have presented NATA, the Nearly Always Truthful Approximations equation of
state: a self-contained chemical-picture stellar EOS for
hydrogen-helium-carbon-oxygen mixtures. NATA carries the bound species
explicitly and returns the pressure, internal energy, and entropy from one
continuous set of expressions across the density-temperature plane. It is not a
committee of stitched tables. Consequently, it has no internal-energy step at
fixed density where one component EOS hands off to another, and a hydrodynamics
code that inverts $u$ for temperature sees no spurious temperature jump.
Across the validated part of the plane, and outside the explicitly regularised cold dense corner and the bounded C/O pressure-ionisation region identified in \S\ref{sec:nata_eos}, the derived quantities $\Gamma_1$, $\nabla_{\rm ad}$, and the sound speed remain smooth through the ionisation, dissociation, and degeneracy fronts.
This is the property required by SPH merger and common envelope calculations, where the flow does not politely stop at EOS
boundaries and ask which table should be trusted next. The applicability of the
EOS by region is summarised in Table~\ref{tab:reliability}: the smoothness and
derivative properties hold across the validated plane,
while the cold dense strongly coupled corner is returned as continuous but
numerically regularised, and the hot rarefied pair wedge is diagnostic only
because pairs are not coupled to the composition solve.

Across the regions where the reference equations of state are used within their domains, NATA agrees at the percent level.  
For the C/O mixture the pressure tracks Skye to better than $0.05\%$ over $95$ per cent of the fully ionised plane, and once the constant fully ionised Skye energy offset of $1.31\times10^{15}$ erg g$^{-1}$ is removed the internal energy agrees with Skye to a median of $0.01\%$ and a ninety-ninth percentile of $0.4\%$;
for the stellar mixture, on the fully ionised reliable plane, the median is $0.05\%$ and the maximum about $1\%$.
 In the dilute molecular limit NATA reproduces the independent ideal
$\mathrm{H_2}\leftrightarrow 2\mathrm H$ balance to better than $0.1\%$ in both
molecular fraction and pressure.  The remaining structures in the comparison
maps, including the partial-ionisation lobes and the cool neutral ridge, are not
single missing constants or erroneous species data; they are the expected
signatures of comparing a chemical-picture EOS with physical-picture, ab initio,
or blended references that place the same fronts differently.

NATA is distributed as a self-contained Fortran module with no external
dependencies and is called through a lightweight driver.  It runs for arbitrary
hydrogen-helium-carbon-oxygen mixtures. In \texttt{gs98} mode the metal mass is partitioned onto the carbon and oxygen proxies using a GS98-like abundance pattern of nineteen elements between carbon and nickel \citep{GrevesseSauval1998}, from which the mean molecular weight then follows, and the user may substitute their own element fractions.   The
code produces the standard $(\log\rho,\log T)$ table and, by inversion at fixed
density, a companion $(\log\rho,\log u)$ table for hydrodynamics codes that
advance internal energy.  Both tables carry temperature, mean molecular weight, pressure, entropy, and, on request, thermodynamic derivatives.  
Because $u(\rho,T)$ is continuous and rises monotonically with temperature
through the ionisation, dissociation, and recombination fronts, the inversion is
single valued there by construction, and over this validated plane the
$(\rho,u)$ table inherits the smoothness of the direct table.  In the strongly
coupled cold dense corner, heavy numerical cancellation can leave isolated cells
without a reliable direct inversion; these are filled only along the same
neighbouring thermodynamic branch.  The resulting $T(\rho,u)$ remains single
valued, but this corner should be regarded as numerically regularised rather than
independently validated.

The EOS has been compared against three independent classes of reference: the blended MESA EOS in its OPAL, SCVH, FreeEOS, HELM, and Skye regions; the CMS19 and CD21 ab initio hydrogen-helium tables; and reference-independent analytic limits, including the degenerate relativistic electron gas and the gas-plus-radiation adiabat of \citet{Chandrasekhar1939}, and the
one-component-plasma Coulomb fit of \citet{PotekhinChabrier2000}.
Where each reference is applied within its own domain, including HELM where the gas is genuinely fully ionised and weakly coupled, the comparison is a validation and NATA agrees at the percent level; where the MESA blend instead falls back outside that domain, such as the HELM and ideal regions returned for cool low density gas that is physically neutral or molecular, the comparison validates nothing about NATA and instead exposes the reference fallback.

The comparisons were used as diagnostics to identify failed approximations.
Several initial treatments were replaced when the comparison maps showed where they failed: the rigid-rotor/harmonic-oscillator $\mathrm{H_2}$ partition function was replaced by the direct rovibrational level sum of \citet{BarklemCollet2016};
the pressure-ionisation closure was replaced by the Hummer-Mihalas occupation probability \citep{HummerMihalas1988,Mihalas1988} with a Yukawa-rescaled Holtsmark microfield \citep{Holtsmark1919}.
The non-relativistic half-integer Fermi-Dirac functions were moved from a single fugacity series, used outside its convergence radius, to a piecewise treatment (a fugacity series where it converges, quadrature through the crossover, and a Sommerfeld expansion when degenerate), and the relativistic electron pressure and energy are evaluated by direct quadrature with the exact \citet{Chandrasekhar1939} zero-temperature limit substituted at extreme degeneracy. The electron EOS was
promoted to the full relativistic finite-temperature form;
the saturated Wigner-Kirkwood ion correction was replaced by the
quantum-liquid fit of \citet{BaikoChugunov2022}, with the liquid OCP energy and
pressure retained and only the entropy regularised across the melting window;
and the C/O ionisation ladder and minor
molecular ions were completed.
These are not fitted corrections.  No comparison-dependent parameter is
adjusted to force agreement with any reference table.

Equally important, not every plausible addition survived.  The standard was not
whether a term sounded physical, but whether it improved the comparisons while
preserving the smoothness and numerical stability required for hydrodynamics.  An
adjustment of the $\mathrm{H_2}$ binding-energy zero point was rolled
back after it drove the cold molecular energy in the wrong direction, confirming
that the original molecular reference zero was the correct one.  The omission is
therefore based on a direct test, not an oversight.

NATA is intended for the full density-temperature range required in mergers of
normal stars: from the hot, fully ionised interior, through the partial-ionisation
and recombination fronts, to the cool, partially molecular envelope.  It can also
be used wherever a transparent chemical-picture mixture is the desired fallback.
It is not a replacement for OPAL, SCVH, or the ab initio hydrogen-helium EOSs in
the regimes for which those references were designed; they contain non-ideal
physics that NATA deliberately does not attempt to reproduce.  The point is
sharper and more practical.  In the cool and low-density corners where a blended
EOS stretches its components beyond their intended domains, 
the MESA blend can
return a component outside the regime it was built for, such as HELM with a fully ionised mean molecular weight where the physical gas would be neutral or molecular.
In those corners NATA returns the continuous chemical-equilibrium state and is
therefore a sensible drop-in replacement for a component used outside its domain, not a rival
to the reference EOS that is appropriate elsewhere.  This regime is not merely a
numerical corner case: the extended envelopes of asymptotic giant branch stars can
enter the same cool, low-density, partially molecular conditions, where a blended
table is most vulnerable to being asked for an answer outside the natural range of
one of its components. 
A continuous chemical-picture EOS is therefore useful not only for stellar mergers and common-envelope calculations, but also for the envelopes of evolved stars.

Used in this way, NATA serves two purposes.  It provides the smooth
interior-to-envelope EOS needed for stellar-merger and common envelope
hydrodynamics, and it provides a safe default in precisely those regions where a
standard blended table is least reliable.  The source code, input files, and the
comparison tables and scripts used in this work are publicly available at
Zenodo, \dataset[10.5281/zenodo.20914774]{https://doi.org/10.5281/zenodo.20914774}.

\begin{acknowledgments}
  N.I. acknowledges funding from NSERC under Discovery grant No. RGPIN-2025-05603. The author acknowledges the use of OpenAI's ChatGPT and Anthropic's Claude as aids for language editing, LaTeX troubleshooting, and code refactoring and manuscript-code consistency checking during preparation of this manuscript.
These tools were not used to generate
scientific results, data, or conclusions.
All scientific choices, derivations, code, interpretations, citations,
and the final text are the author's own work.
AI assistance was used only in a supporting role, and the author remains fully responsible for the content of the paper.
\end{acknowledgments}

\bibliographystyle{aasjournal}
\bibliography{references}

\end{document}